\UseRawInputEncoding

\documentclass[aps,prl,amsmath,amssymb,floatfix,reprint,citeautoscript,noeprint,superscriptaddress,twocolumn]{revtex4-2}
\usepackage{xr}

\usepackage{dsfont}
\usepackage{lipsum} 
\usepackage{bibentry}
\usepackage[english]{babel}
\usepackage[dvipsnames]{xcolor}
\usepackage{graphicx}
\usepackage[caption=false]{subfig} 
\usepackage{amsmath,amssymb,bm}
\usepackage[version=3]{mhchem}
\usepackage{verbatim}
\usepackage{multirow}
\usepackage{dcolumn}
\usepackage{float}
\usepackage{ragged2e}
\usepackage{nicefrac}
\usepackage{siunitx}
\usepackage{booktabs}
\usepackage{chemformula}
\usepackage{wrapfig}
\usepackage{enumitem}  
\usepackage{transparent}
\usepackage[colorlinks,allcolors=black,citecolor=blue,urlcolor=blue]{hyperref}
\usepackage{tcolorbox}
\usepackage{relsize}

\newcommand{\etal}{\emph{et al.}}

\usepackage{sansmathfonts}
\makeatletter
\long\def\@makecaption#1#2{%
  \vskip\abovecaptionskip
  \sffamily %
  \small    %
  \begingroup
    \leftskip=0pt plus 1fil
    \rightskip=0pt plus 1fil
    \parfillskip=0pt
    \justifying %
    \noindent
    \bfseries #1: \normalfont #2\par
  \endgroup
  \vskip\belowcaptionskip
}
\makeatother


\begin{document}

\def\mytitle{
Towards Routine Condensed Phase Simulations with Delta-Learned Coupled Cluster Accuracy: Application to Liquid Water
}
\title{\mytitle}
\author{Niamh O'Neill}%
\email{nco24@cam.ac.uk}
\affiliation{%
Yusuf Hamied Department of Chemistry, University of Cambridge, Lensfield Road, Cambridge, CB2 1EW, UK
}
\affiliation{%
Cavendish Laboratory, Department of Physics, University of Cambridge, Cambridge, CB3 0HE, UK
}
\affiliation{%
Lennard-Jones Centre, University of Cambridge, Trinity Ln, Cambridge, CB2 1TN, UK
}
\author{Benjamin X. Shi}%
\email{mail@benjaminshi.com}
\affiliation{Initiative for Computational Catalysis, Flatiron Institute, 160 5th Avenue, New York, NY 10010}

\author{William J. Baldwin}%
\affiliation{%
Lennard-Jones Centre, University of Cambridge, Trinity Ln, Cambridge, CB2 1TN, UK
}
\affiliation{%
Department of Engineering, University of Cambridge, Cambridge, CB3 0HE, UK
}

\author{William C. Witt}%
\affiliation{%
Harvard John A. Paulson School of Engineering and Applied Sciences, Harvard University, Cambridge, MA, USA
}
\author{G\'abor Cs\'anyi}
\affiliation{%
Lennard-Jones Centre, University of Cambridge, Trinity Ln, Cambridge, CB2 1TN, UK
}
\affiliation{%
Department of Engineering, University of Cambridge, Cambridge, CB3 0HE, UK
}
\author{Julian D. Gale}%
\affiliation{%
School of Molecular and Life Sciences, Curtin
University, PO Box U1987, Perth, Western Australia 6845, Australia
}
\author{Angelos Michaelides}%
\affiliation{%
Yusuf Hamied Department of Chemistry, University of Cambridge, Lensfield Road, Cambridge, CB2 1EW, UK
}
\affiliation{%
Lennard-Jones Centre, University of Cambridge, Trinity Ln, Cambridge, CB2 1TN, UK
}
\author{Christoph Schran}%
\email{cs2121@cam.ac.uk}
\affiliation{%
Cavendish Laboratory, Department of Physics, University of Cambridge, Cambridge, CB3 0HE, UK
}
\affiliation{%
Lennard-Jones Centre, University of Cambridge, Trinity Ln, Cambridge, CB2 1TN, UK
}

\begin{abstract}
Simulating liquid water to an accuracy that matches its wealth of available experimental data requires both precise electronic structure methods and reliable sampling of nuclear (quantum) motion.
This is challenging because applying the electronic structure method of choice -- coupled cluster theory with single, double and perturbative triple excitations [CCSD(T)] -- to condensed phase systems is currently limited by its computational cost and complexity. 
Recent \textit{tour-de-force} efforts have demonstrated that this accuracy can indeed bring simulated liquid water into close agreement with experiment using machine learning potentials (MLPs).
However, achieving this remains far from routine, requiring large datasets and significant computational cost.
In this work, we introduce a practical approach that combines developments in MLPs with local correlation approximations to enable routine CCSD(T)-level simulations of liquid water. 
When combined with nuclear quantum effects, we achieve agreement to experiments for structural and transport properties.
Importantly, the approach also handles constant pressure simulations, enabling MLP-based CCSD(T) models to predict isothermal-isobaric bulk properties, such as water's density maximum in close agreement with experiment.
Encompassing tests across electronic structure, datasets and MLP architecture, this work provides a practical blueprint towards routinely developing CCSD(T)-based MLPs for the condensed phase.
\end{abstract}

{\maketitle}

\section{Introduction}

Water is a widely studied system of fundamental scientific and technological importance.
Its extensive body of experimental data makes it a fertile testing ground for evaluating new developments to atomistic simulation methods.
More broadly, it is the prototypical `condensed phase' system, serving as a bridge between gas-phase molecules and solid-state compounds.
For this system, achieving agreement with experiments requires, first, an accurate electronic structure method capable of capturing the delicate hydrogen bond network governing its potential energy surface (PES) \cite{gillanPerspectiveHowGood2016}, and second, sufficient sampling of the nuclear quantum motion on this PES \cite{ceriottiNuclearQuantumEffects2016}.
Generally, satisfying both criteria simultaneously is challenging, often requiring a trade-off between the accuracy of the electronic structure method and computational efficiency for reliable sampling.

In the recent decade, the rise of machine learning potentials (MLPs) has helped to alleviate the aforementioned cost-accuracy tradeoff, serving as efficient surrogate models trained to reproduce electronic structure methods \cite{behlerPerspectiveMachineLearning2016,bartokMachineLearningUnifies2017,thiemannIntroductionMachineLearning2024,kangLargeScaleAtomicSimulation2020}.
To date, density functional theory (DFT) has been the workhorse for generating reference training data for MLPs, owing to its mature implementation of periodic boundary conditions (PBC) -- the most natural means of describing condensed-phase systems -- and availability of energy gradients (forces) that make it cost-effective for generating suitable datasets for training condensed-phase MLPs.
The utility of DFT-trained MLPs for water has been demonstrated in several recent works, enabling new insight into its condensed phase thermodynamics, particularly its phase diagram \cite{reinhardtQuantummechanicalExplorationPhase2021, zhangPhaseDiagramDeep2021} and how its intricate hydrogen bond network governs its unique properties \cite{morawietzHowVanWaals2016}.
However, there are systematic and clear failings in DFT that can prevent the reliable reproduction of experiments.
For the case of liquid water, there is a common over-structuring of the radial distribution function (RDF) \cite{gillanPerspectiveHowGood2016, monterodehijesDensityIsobarWater2024,chenInitioTheoryModeling2017}.
More broadly, there have been ongoing challenges in predicting the position of the density maximum and the density ordering between ice and liquid water with DFT \cite{gaidukDensityCompressibilityLiquid2015, chenInitioTheoryModeling2017} as well as the general phase diagram of water~\cite{boreRealisticPhaseDiagram2023,sciortinoConstraintsLocationLiquid2025}.

To faithfully describe the PES, the method of choice is the so-called `gold-standard' coupled cluster theory with single, double and perturbative triple excitations [CCSD(T)].
This method has been shown to accurately reproduce experimental results for small gas-phase molecules~\cite{kartonW4TheoryComputational2006}, and more recently, for surfaces \cite{shiManyBodyMethodsSurface2023,shiAccurateEfficientFramework2025b,schaferLocalEmbeddingCoupled2021a,carboneCOAdsorptionPt1112024,yeAdsorptionVibrationalSpectroscopy2024} and materials~\cite{yePeriodicLocalCoupledCluster2024a,yangInitioDeterminationCrystalline2014a}.
However, using it to train condensed-phase MLPs presents a significant challenge, due to: (1) its prohibitive cost (formally scaling as $N^7$ with $N$ electrons), (2) PBC implementations of CCSD(T) remain underdeveloped, and (3) gradients of the CCSD(T) PES are difficult to obtain.
Nevertheless, recent work has highlighted promising approaches to bypass these bottlenecks for liquid water using MLPs.
Daru and co-workers used a $\Delta$-learning strategy where the difference between CCSD(T) with the domain-based local pair natural orbital (DLPNO) approximation~\cite{riplingerEfficientLinearScaling2013} and a more affordable level of theory is fitted, showing that energies (without gradients) are sufficient to train on gas-phase clusters~\cite{daruCoupledClusterMolecular2022}.
This $\Delta$-MLP is added onto a baseline MLP trained to the lower-level of theory (such as DLPNO-accelerated second-order M\o{}ller-Plesset perturbation theory (MP2) and periodic DFT \cite{meszarosShortRangeDMachineLearning2025a}) to reach the final desired level of accuracy.
Separately, Chen \etal{} showed that combining new PBC implementations of CCSD(T) --- made more efficient with the frozen natural orbital approximation --- with transfer learning enables data-efficient learning from small computationally tractable unit cells \cite{chenDataEfficientMachineLearning2023}.

The aforementioned works have highlighted the utility and promise of CCSD(T)-level MLPs, reaching agreement with experiment across select properties of liquid water.
Such agreement has also previously been achieved by alternative models based upon the many-body expansion (MBE), such as MB-pol \cite{palosCurrentStatusMBpol2024} or q-AQUA-pol \cite{quInterfacingQAQUAPolarizable2023}, which have achieved great insight into water's complex phase behavior \cite{sciortinoConstraintsLocationLiquid2025} .
One of the medium-term goals of MLPs is to utilize their general applicability and easily derivable nature for diverse systems, while retaining the same quality realized by MBE approaches~\cite{boreRealisticPhaseDiagram2023,sciortinoConstraintsLocationLiquid2025,bowmanPerspectiveMarking202025}.
This is crucial in order to improve over the MBE which --- despite its successes --- suffers from the permutational growth in body terms, steep rise in complexity with number of species, fixed topology and bespoke development process.

There are open questions that need to be addressed to enable routine CCSD(T)-level MLPs for condensed phase simulations.
Firstly, to date, no CCSD(T) MLP model of liquid water has been applied at constant pressure, which is important to fully resolve the isothermal-isobaric properties of a system.
For liquid water, this is necessary to predict the equilibrium density of water, which exhibits a well-known (anomalous) density maximum as a function of temperature.
Accurately describing density fluctuations is especially challenging when models are trained only on clusters, as missing long-range interactions from utilizing short-range MLPs tend to increase the predicted density \cite{zaverkinPredictingPropertiesPeriodic2022, kovacsMACEOFFShortRangeTransferable2025}.
This has been explored at the DFT level by Zaverkin \etal{}~\cite{zaverkinPredictingPropertiesPeriodic2022}, which showed MLPs trained directly on water clusters can predict errors in the density of up to 10~\% w.r.t.\ a model trained on periodic data.
New approaches are needed that can accurately predict the density to within $1-3\,$\% from cluster data.

The previous discussed CCSD(T) MLP models of water have heralded the start of CCSD(T) MLPs for the condensed phase.
However, they were also \textit{tour de force} efforts, requiring significant computational investment, thereby limiting the efficient and routine development of such models.
For example, the study by Daru \etal{} required roughly 3,000 and 13,000 clusters of 64 waters at the CCSD(T) and MP2 level, respectively, amounting to ${\sim}3.7\,$million CPU hours (CPUh) to compute, while the computational cost of periodic CCSD(T) limited Chen \etal{} to small 16-water molecule boxes.
To lower these costs, it will be important to benchmark the effect of electronic structure parameters -- namely the basis set and the thresholds on the local approximations to CCSD(T) -- on condensed phase properties.
Such benchmarks until now, have typically focused on gas-phase energetics, which are difficult to connect to thermodynamic observables.
This was briefly explored by Chen \etal{} for two basis sets (TZV2P from VandeVondele and Hutter~\cite{vandevondeleGaussianBasisSets2007} and cc-pVQZ from Dunning~\cite{petersonAccurateCorrelationConsistent2002}), which reported marked changes in the OH stretch peak.
However, a systematic study, particularly involving the typical hierarchy of basis sets from double (DZ), to triple (TZ) and quadruple (QZ) $\zeta$ in size, such as within the correlation consistent (cc) basis set family, has not been performed, as well as the effect of two-point complete basis set extrapolations~\cite{fellerEffectivenessCCSDTComplete2011} commonly employed to enable smaller basis sets.

In this work, we present a practical and efficient blueprint for developing MLP models at the CCSD(T) level of theory, bringing together the many advances described above towards the accurate simulation of the thermodynamics of liquid water.
We tackle the three challenges described in the previous paragraph as follows:
(1) We build on the aforementioned $\Delta$ learning strategies for CCSD(T) MLPs, to now enable simulations at constant pressure, allowing the density properties of water to be probed, specifically the density isobar of water.
(2) We present a computationally efficient approach to achieve CCSD(T) level MLPs, enabled by improved data efficiency by using the MACE MLP approach as well as cheaper electronic structure settings, leading to over two orders of magnitude cheaper costs.
We have validated (2) by benchmarking the (3) effect of the electronic structure parameters directly on experimental thermodynamic properties of interest.

\begin{figure*}[t]
    \includegraphics[width=1.0\textwidth]{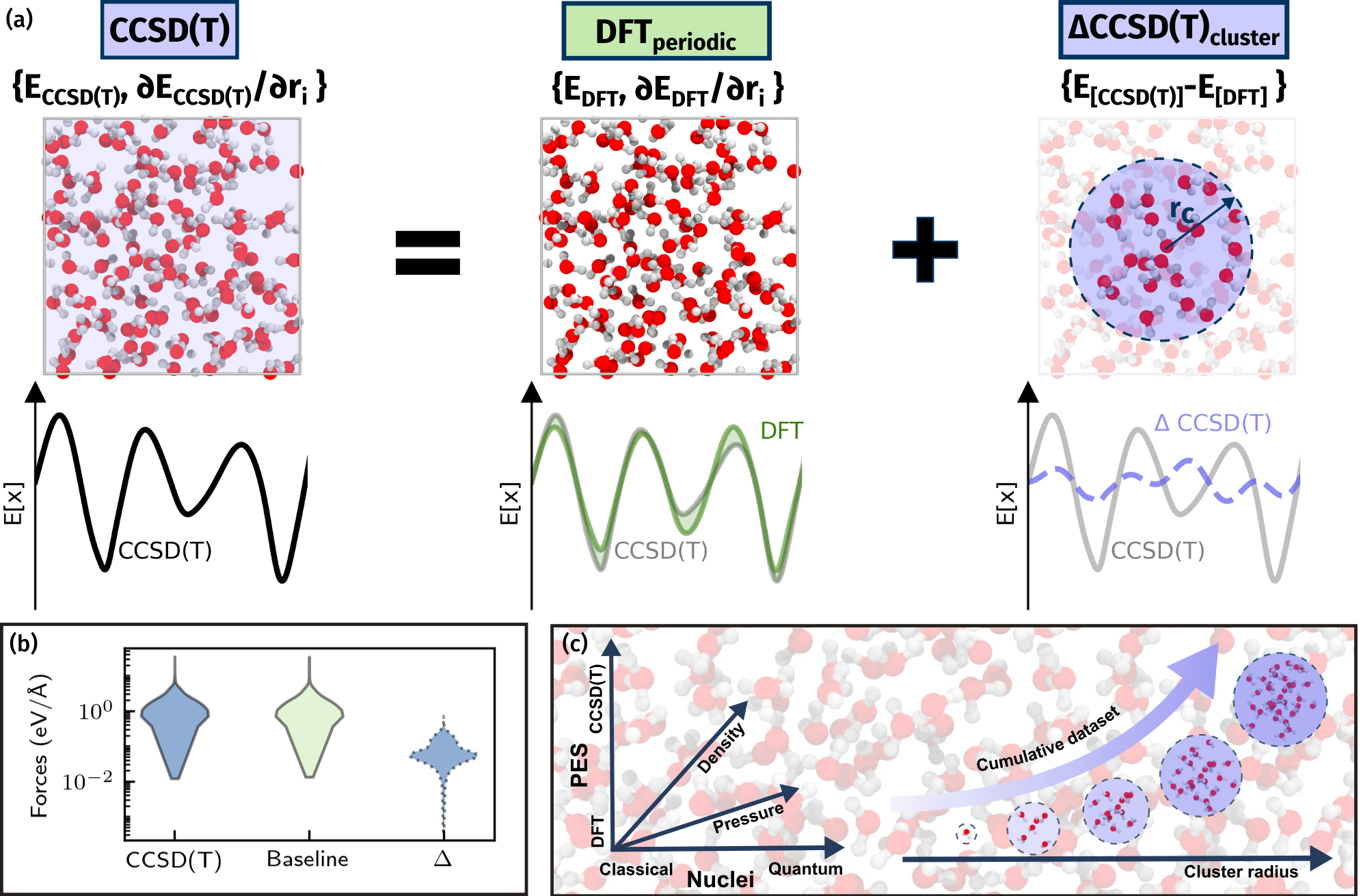}
    \caption{%
    \textbf{Schematic of the approach to reach CCSD(T) accuracy for liquid water.}
    (a) A periodic DFT MLP model is corrected via a  $\Delta$-MLP model trained on the energy difference of gas phase clusters between DFT and CCSD(T). The relative magnitudes of the PESs to be learned are illustrated schematically below each snapshot. (b) Violin plots of the predicted force distributions of the periodic dataset by the CCSD(T), baseline and $\Delta$-MLP models. (c) Schematic of the conditions sampled for generating the periodic baseline and $\Delta$-MLP datasets. The $\Delta$-MLP model dataset contains cumulatively increasing cluster radii.}
    \label{fig:schematic}
\end{figure*}

\section{Reaching a converged CCSD(T) Delta-MLP}\label{sec:approach}

In this section, we propose several new developments to the $\Delta$-learning framework of Daru and co-workers \cite{daruCoupledClusterMolecular2022, meszarosShortRangeDMachineLearning2025a}.
Our approach generates a cost-efficient training set, enabling routine development of CCSD(T)-level MLPs that can handle constant pressure condensed-phase simulations, which we will later leverage to predict the density of liquid water.
The upper panel of Figure~\ref{fig:schematic} summarizes the $\Delta$-learning approach used within this work to develop the CCSD(T) MLPs for liquid water.
Starting from a `baseline' MLP trained to periodic DFT data, a further $\Delta$-MLP is fitted to elevate the PES to the CCSD(T) level.
This $\Delta$-MLP is trained on energy differences (without gradients) between the baseline DFT and CCSD(T) from gas phase clusters extracted from equilibrium molecular dynamics simulations.
We further exploit various local CCSD(T) approximations, such as the aforementioned DLPNO as well as the local natural orbital (LNO)~\cite{nagyOptimizationLinearScalingLocal2018,gyevi-nagyIntegralDirectParallelImplementation2020}  approximation, to enable tractable calculations of much larger clusters than feasible with canonical CCSD(T).
The final CCSD(T) MLP is the sum of (energies and forces) predicted by the baseline and $\Delta$-MLPs, which is used to subsequently predict structural and dynamical properties of liquid water.
Panel (b) of Figure \ref{fig:schematic} quantitatively demonstrates the reduced complexity of the fitting task for the $\Delta$-MLP model relative to the baseline. 
Here we compare the distribution of predicted forces on a set of periodic snapshots by the baseline, CCSD(T) and $\Delta$-MLP models, where the predicted forces are an average of two orders of magnitude lower for the $\Delta$-MLP model.

As demonstrated by Zaverkin~\etal{}~\cite{zaverkinPredictingPropertiesPeriodic2022} and Kovacs~\etal{}~\cite{kovacsMACEOFFShortRangeTransferable2025}, up until now, MLPs trained directly on clusters cannot accurately simulate under constant pressure, resulting in an overestimation of the density.
A key observation of this current work is that if the baseline MLP trained to periodic DFT data can reliably perform constant pressure simulations, the resulting CCSD(T) MLP can yield accurate constant pressure simulations with a $\Delta$-MLP trained to cluster data.
While the use of a baseline trained to periodic MBE-based potential (fitted to DFT) was performed by Meszaros \etal{}~\cite{meszarosShortRangeDMachineLearning2025a}, their work did not highlight its utility for constant pressure simulations.

Panel (c) highlights the dataset considerations in this work in order to ensure reliable simulations under constant pressure.
The baseline dataset includes configurations sampled across a wide range of thermodynamic conditions, encompassing various pressures and densities.
Nuclei described both classically and including nuclear quantum effects (NQEs) via path integral molecular dynamics (PIMD) simulations are also sampled, as well as configurations from both the baseline and CCSD(T) simulations.
This diverse pool of structures is also directly used to generate the clusters used for the $\Delta$-MLP dataset.
While previous works have focused on generating datasets with clusters of a fixed number of water molecules, we show in Section \ref{si-sec:data_efficiency} of the SI and schematically in panel (c) of Figure \ref{fig:schematic} that datasets containing clusters of cumulatively increasing sizes allows for significantly fewer large clusters, which take up the significant contribution to the CPU costs of the training set.

\begin{figure*}[ht]
    \includegraphics[width=1.0\textwidth]{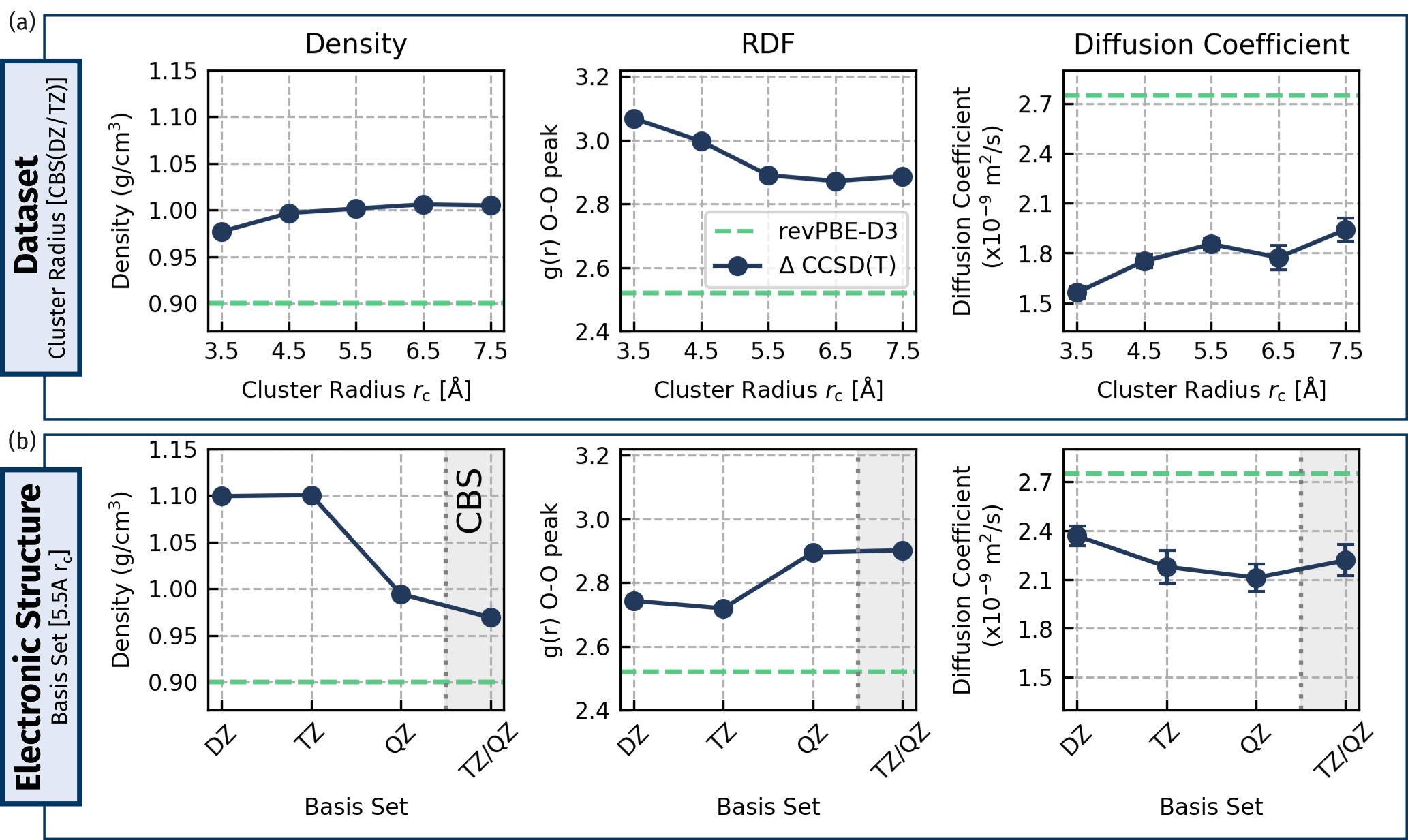}
    \caption{%
    \textbf{A `self-consistent' approach to determine convergence of the $\Delta$-MLP dataset.}
    We converge the (a) cluster size and (b) CCSD(T) basis set for the structural and dynamical properties of bulk liquid water with classical nuclei (specifically the density, first O-O RDF peak height and diffusion coefficient). The cluster radius datasets are a cumulative sum of all smaller radii datasets up to a given cutoff $r_c$. The shaded grey region in the lower plots indicates complete basis set (CBS) extrapolation. Here the periodic baseline is revPBE-D3, whose results are also shown for reference with a green dashed line. The cluster size tests are done at the CBS(DZ/TZ) level of theory, while basis sets tests are done with the 5.5~\AA ~dataset.}
    \label{fig:convergence}
\end{figure*}

A particular advantage of our proposed approach is that it provides a `self-consistent' means to validate the accuracy of the $\Delta$-MLP.
The cluster dataset is generated by progressively incorporating clusters of increasing sizes until the properties of interest converge.
It assumes that the condensed-phase properties can be learned in the limit of large cluster sizes and we have demonstrated that the differences between two DFT levels can be learned with such an approach in Section \ref{si-sec:dft2dft} of the SI (where in this case we can compare to models trained on periodic data).
This has also been demonstrated by Meszaros \etal{}~\cite{meszarosShortRangeDMachineLearning2025a} between two MBE-based potentials.
In panel (a) of Figure ~\ref{fig:convergence}, we plot the convergence of cumulative datasets containing cluster sizes up to a radius ($r_c$) of $7.5\,$\AA{}, going up in intervals of $1.0\,$\AA{} starting from $2.5\,$\AA{}, with ${\sim}1800$ structures at each cluster size.
We directly compare how the resulting CCSD(T) MLPs converge key thermodynamic observables such as the density, self-diffusion coefficient and radial distribution function (RDF) -- specifically the first peak of the O--O RDF -- at ambient conditions.
There is a well-controlled convergence for all three properties with maximum cluster size and we find that a radius of $5.5\,$\AA{} -- corresponding to ${\sim}23$ water molecules -- is required to converge all three properties, reproducing the density to within 0.1\% of the $7.5\,$\AA{} result, the diffusion constant to within 4.5\%, and the O-O RDF peak to within $0.01$.
These tests use a revPBE-D3 DFT baseline, which severely underestimates the density by $10\,$\% and the $\Delta$-MLP corrects the DFT to CCSD(T) level and close to experiment ($0.997\,$g/cm$^3$ at $298\,$K) while also bringing the other observables in a direction towards experiment upon inclusion of nuclear quantum effects, as will be shown in the next section.

The dataset convergence tests were performed at the CBS(DZ/TZ) level of theory because larger basis sets at the CCSD(T) level for the $6.5\,$ and $7.5\,$\AA{} clusters have a prohibitively large computational cost.
Regardless, we expect these observations to persist for larger basis sets, given the clear convergence observed in Figure ~\ref{fig:convergence}.
Towards this end, with the converged dataset (up to $5.5\,$\AA{}), we have tested basis sets increasing in size from double zeta (DZ) to triple zeta (TZ) and quadruple zeta (QZ) for jul-cc-pV$X$Z basis family~\cite{papajakPerspectivesBasisSets2011}, as well as the reliability of two-point extrapolation to the complete basis set (CBS) limit in the bottom panel of Figure~\ref{fig:convergence}.
Our tests indicate that the QZ basis set is well converged for the thermodynamic properties studied here.
The density is highly sensitive to the basis set, with a decrease of 10\% going from TZ to QZ.
While it was too costly to go towards larger basis sets, we can approximate these with CBS extrapolations~\cite{neeseRevisitingAtomicNatural2011}.
In particular, QZ is in close agreement with extrapolated CBS(TZ/QZ) for all three properties, giving us confidence that it has converged, and we utilize the QZ basis set in our final models.

\begin{figure*}[t]
\centering
\includegraphics[width=1.0\textwidth]{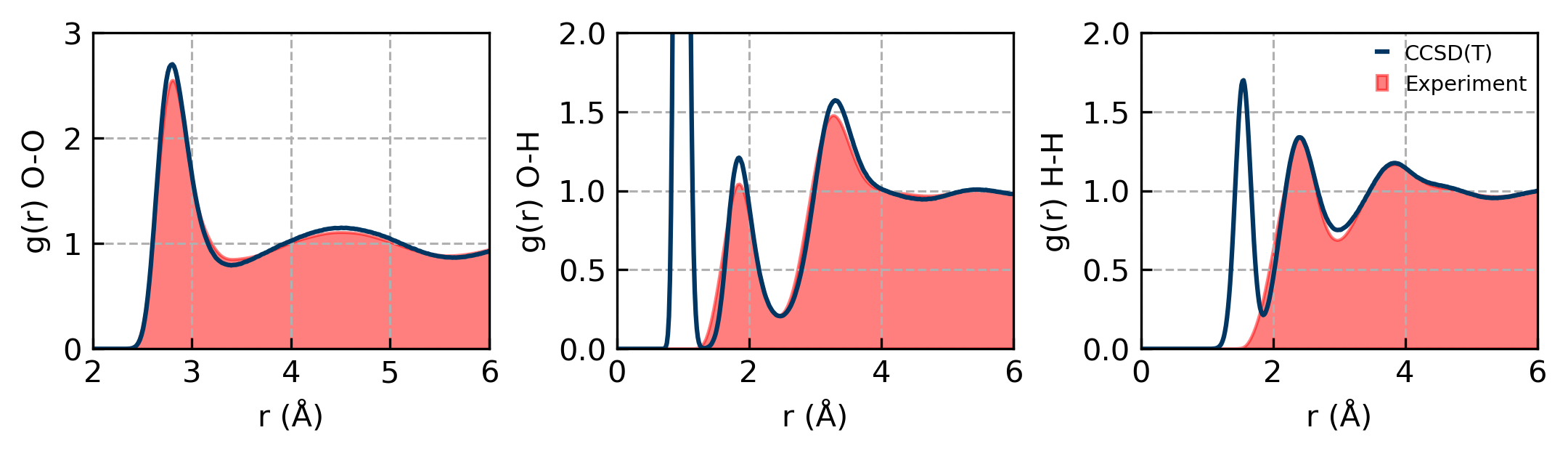}
\caption{
\textbf{Structure of CCSD(T) water from RPMD simulations.}
RDFs are computed at the equillibrium CCSD(T) density of 0.989 g/cm$^3$ that has been obtained from NPT simulations. The experimental O-O RDF is taken from a temperature interpolation of X-ray diffraction data from \cite{skinnerStructureWaterCompressibility2014} to 298K by Daru et al \cite{daruCoupledClusterMolecular2022}. O-H and H-H RDFs are from total scattering data from Ref. \cite{soperRadialDistributionFunctions2013}. }
\label{fig:results}

\end{figure*}

The final (dubbed `full') dataset we use is at the jul-cc-pVQZ level and consists of ${\sim}7000$ structures comprising of ${\sim}1800$ clusters each that are 2.5, 3.5, 4.5 and $5.5\,$\AA{} in radius.
While the convergence tests shown here have been performed with revPBE-D3 (with zero damping) as the DFT baseline, any choice of DFT baseline is possible and we show that the resulting properties are in agreement using both PBE-D3 (with zero damping) and r$^2$SCAN as baselines in Section \ref{si-sec:dft2ccsdt} of the SI.
For our final simulations in subsequent sections, we have opted to use r$^2$SCAN as the baseline model as its difference with CCSD(T) is the lowest -- corresponding to lower average predicted forces from the $\Delta$-MLP model discussed in Section  \ref{si-sec:dft2ccsdt} of the SI.

\section{Structure and Dynamics of CCSD(T) water}\label{sec:results}

Figure~\ref{fig:results} and Table \ref{tab:results} compares the predictions of the final CCSD(T) model – incorporating NQEs via (thermostatted) ring polymer molecular dynamics ((T-)RPMD) for dynamical and structural properties against experimental data.
We achieve good agreement to experimental data for all properties, spanning both structural properties such as the RDF and density, as well as transport properties -- specifically the diffusion coefficient.
We compare the O-O, O-H and H-H RDFs, with the CCSD(T) predictions overlapping experiments across all three properties in Figure \ref{fig:results}.
The diffusion coefficient is predicted at 298 K to be 0.22 $\pm$ 0.01~ \AA$^2$/ps, within the error bars of experiment and in addition, the predicted density of 0.989 $\pm$ 0.003 g/cm$^3$ at 298 K is less than 1\% from the predicted experimental value at 298 K of 0.997 g/cm$^3$ as shown in Table \ref{tab:results}.
Given that our CCSD(T) MLP model is able to handle constant pressure simulations, we have computed both the RDFs and diffusion coefficient at the appropriate CCSD(T) equilibrium density; this is particularly valuable because not all condensed phase systems have available experimental data of the density.

The good agreement with experiment has been enabled by the combination of not only a high-quality CCSD(T) MLP, but also the incorporation of NQEs.
In Section \ref{si-sec:classical} of the SI, we show the predictions arising from classical molecular dynamics without NQEs.
We find that neglecting NQEs leads to a over-structuring of the RDFs and a lowering of the diffusion coefficient.
These results are in line with previous work suggesting a roughly 1.15 increase on diffusion from inclusion of NQEs \cite{habershonCompetingQuantumEffects2009}.
On the other hand, the effect of NQEs on the density are negligible - with a slight decrease upon their inclusion in line with previous works \cite{meddersDevelopmentFirstPrinciplesWater2014}.

\begin{table}[h]
\caption{Summary of density and diffusion coefficient from RPMD and T-RPMD simulations respectively at 298 K compared to experiment.
Simulation estimates are corrected for finite size effects and standard errors from 9 independent simulations are reported in parenthesis.}
\begin{tabular}{@{}llll@{}}
\toprule
                 & $\rho ~[\mathrm{g/cm^3}$] & D ~[A$^2$/ps]  \\ \midrule
Experiment       & 0.997                  & 0.23                                     \\
CCSD(T) & 0.989 (0.003)          & 0.22 (0.01)                            
\end{tabular}
\label{tab:results}
\end{table}

\begin{figure}[ht]
    \centering
    \includegraphics[width=1\linewidth]{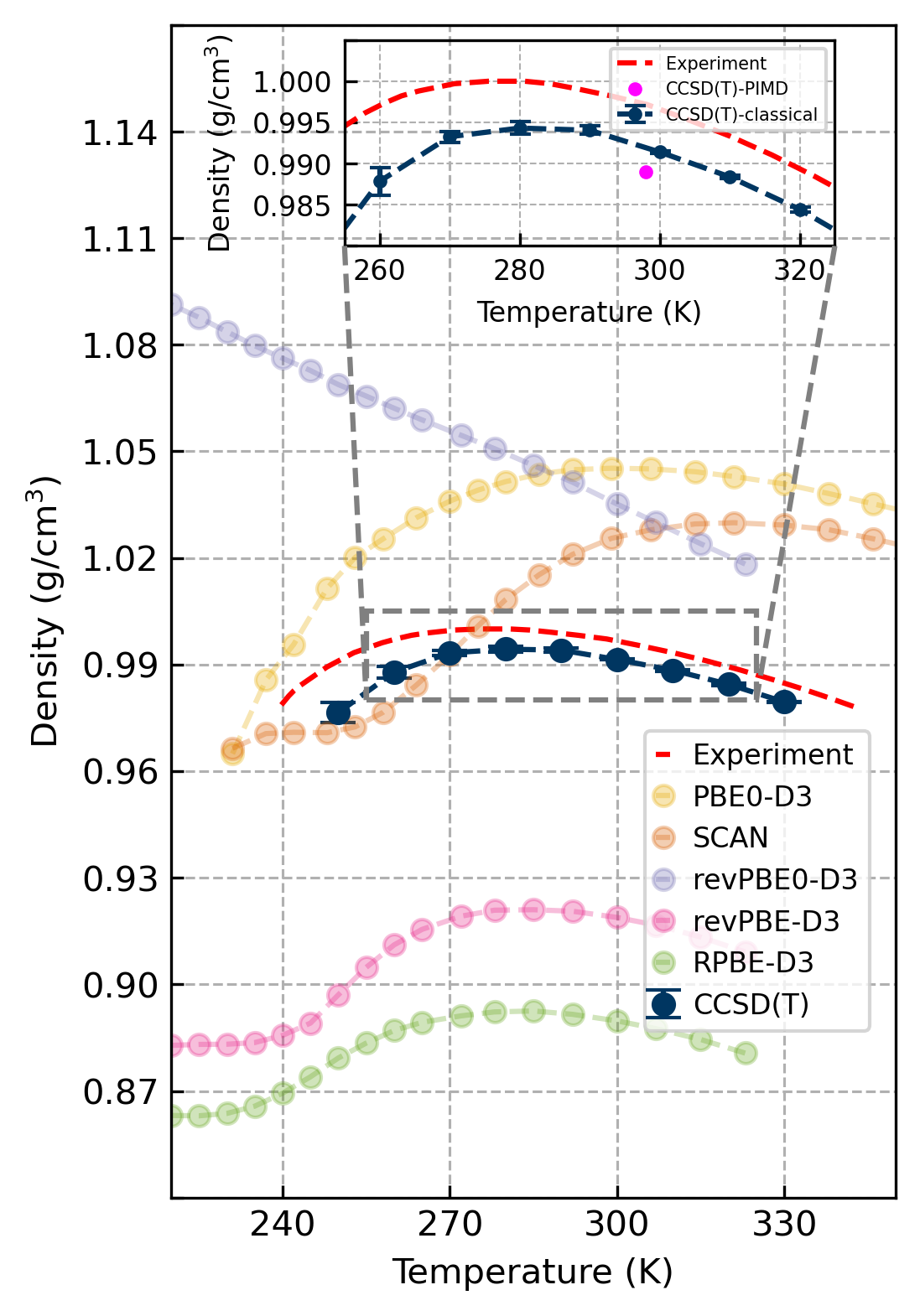}
    \caption{\textbf{Density isobar of liquid water.}
    The main panel compares the density isobar of the $\Delta$ CCSD(T) model to common DFT functionals revPBE-D3, revPBE0-D3, RPBE-D3 and SCAN, as well as experiment. All simulations used classical nuclei. DFT has been reproduced from Ref. \cite{monterodehijesDensityIsobarWater2024}. All D3 shown correspond to the zero-damping variant of dispersion correction from Grimme \etal{}. The inset highlights the experimental and CCSD(T) isobars, as well as showing the slight decrease in the CCSD(T) density upon inclusion of NQEs.}
    \label{fig:isobar}
\end{figure}

The generalizability of the models can be explored by computing the density (and other) properties of water beyond ambient conditions.
We showcase this for the density isobar of water between $250\,$K and $330\,$K in Figure \ref{fig:isobar}.
This is a particularly challenging property for DFT to predict, as recently shown by de Hijes \etal{}~\cite{monterodehijesDensityIsobarWater2024}.
In Figure \ref{fig:isobar}, we compare our CCSD(T) MLP predictions against a selection of common DFT functionals, taken from Ref.\citenum{monterodehijesDensityIsobarWater2024}.
With our CCSD(T) MLP, we achieve agreement to experiments across the entire temperature range to within 1.4\%.
We predict a density maximum at 280 K, which is within 1.1\% of the experimental value of 277 K.

\section{Towards cost effective CCSD(T) MLPs}

\begin{figure*}[ht]
\centering
\includegraphics[width=1.0\textwidth]{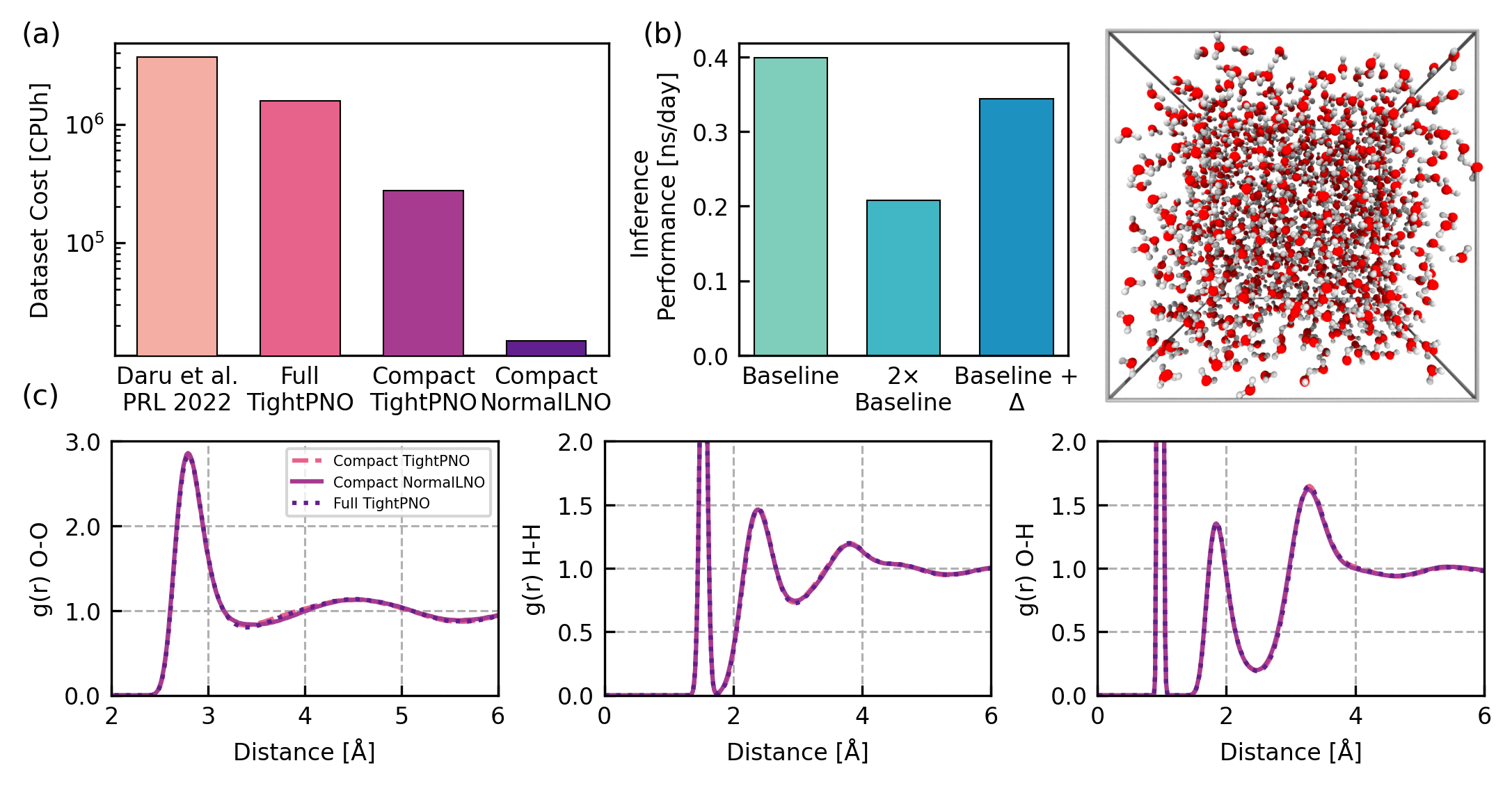}
\caption{
\textbf{Efficient cost of the $\Delta$-learning approach.} Panel (a) shows the computational cost, in CPU hours, required to generate datasets for our method compared to previous CCSD(T) $\Delta$ ML potential studies by Daru et al. \cite{daruCoupledClusterMolecular2022}. The last three bars correspond to: (i) the most conservative model using the full dataset with tight PNO thresholds in the DLPNO-CCSD(T) approximation, (ii) the compact dataset with TightPNO thresholds in DLPNO-CCSD(T), and (iii) the compact dataset with normal LNO thresholds (abbreviated as NormalLNO) in LNO-CCSD(T). Panel (b) shows the inference performance in ns/day (using 1 fs timestep) of 1) the baseline MACE model architecture, 2) the hypothetical sum of 2 baseline architectures, and 3) the baseline and $\Delta$-MLP model summed for a simulation box of 1000 waters shown adjacent. Panel (c) presents the predicted radial distribution functions (RDFs) for the full and compact datasets and conservative and relaxed electronic structure settings shown in Panel (a).
 }
\label{fig:cost}
\end{figure*}

The purpose of the above section was to accurately reach convergence of the CCSD(T) properties without focusing on data-efficiency and cost-efficiency, using a large dataset as well as conservative local CCSD(T) parameters.
As a result, the dataset required roughly $1.6\,$million CPUh to compute, as shown in panel (a) of Figure ~\ref{fig:cost}.
We show in Section~\ref{si-sec:data_efficiency} of the SI that the majority of this cost (87\%) comes from the ${\sim}1800$ of the largest $5.5\,$\AA{} clusters.
However, we find that the number of $5.5\,$\AA{} clusters can be lowered to 100 without affecting this accuracy.
In particular, this was made possible because we used a cumulative dataset (incorporating clusters of smaller sizes) and we find that utilizing \textit{e.g.}, only $5.5\,$\AA{} clusters would require over 1500 clusters to converge all studied thermodynamic observables.
This resulting dataset (containing 5500 clusters and of which only 100 are $5.5\,$\AA{} in radius) -- dubbed `compact dataset' achieves the same accuracy on all studied observables while being more than one order of magnitude cheaper.
We show this for the RDFs in Figure \ref{fig:cost}(c).
Furthermore, without any loss in accuracy for the predictions of the thermodynamic observables, we find that the conservative ``TightPNO'' DLPNO approximation (${\sim}700\,$CPUh for each $5.5\,$\AA{} cluster) can be relaxed to use the LNO approximation with ``normal'' thresholds, resulting in only ${\sim}30\,$CPUh for each $5.5\,$\AA{} cluster.
As shown in Section~\ref{si-sec:cheapest} of the SI, with all of these optimizations, the total cost of the dataset can be lowered to $15,000\,$CPUh, requiring a maximum RAM of $50\,$GB and at a cost that can be achieved within a couple of weeks on a desktop.

In principle, the disadvantage of $\Delta$-learning is an increase in inference time (energy and gradient evaluation) due to the need to evaluate both the baseline and $\Delta$-MLP models.
However, it has been highlighted by Meszaros \etal{}~\cite{meszarosShortRangeDMachineLearning2025a} that the $\Delta$-MLP can be fitted more easily than the baseline, requiring shorter cutoffs.
In Section \ref{si-sec:ml} of the SI, we show that a 
much smaller architecture
compared to the baseline model is sufficient to accurately capture the bulk behavior for the $\Delta$-MLP.
Importantly, this makes the inference cost of the $\Delta$-MLP almost negligible compared to the large MACE baseline model.
Panel (b) of Figure~\ref{fig:cost} shows only a modest reduction of approximately 10\% in inference performance when summing the baseline and $\Delta$-MLP within the final CCSD(T) model compared to the baseline alone -- significantly more efficient than the hypothetical case of summing two full baseline models.
These tests were performed on a simulation box containing 1000 water molecules on 128 AMD EPYC 7742 cores, using the Symmetrix library, a C$++$, Kokkos optimised MACE implementation \cite{wittSymmetrixAvailableHttps2025, kovacsEvaluationMACEForce2023}.
Coupling the CCSD(T) models with such efficient software further underscores the possibilities for large-scale molecular dynamics simulations with CCSD(T) level accuracy.

\section{Discussion}
The level of accuracy demonstrated in this paper for liquid water has been achieved before by previous CCSD(T) MLP models~\cite{chenDataEfficientMachineLearning2023, daruCoupledClusterMolecular2022} as well as models based on the MBE, notably MB-pol and q-AQUA~\cite{yuStatusReportGold2023, quInterfacingQAQUAPolarizable2023}.
Our agreement with experiment mirrors these previous works as shown in Section \ref{si-sec:comparison-literature} of the SI.
The key development in this work lies in providing a practical and generalizable approach for constructing CCSD(T)-level MLPs for condensed phase systems.
Crucially,
our approach enables simulations that can be performed under constant pressure,
where we demonstrate that the models are able to capture the density isobar of water.
In particular, we have shown that it is possible to provide systematic convergence tests to identify key parameters in the training dataset that ensure resulting properties which are converged and can enable efficient development of CCSD(T) level MLPs.
All of the models trained in this work, datasets and specific electronic structure inputs as well as a small code to cut the clusters will be made available upon publication, as a starting point for others to develop CCSD(T) level models. %

However, this current approach has some limitations that can hinder the progress towards the aforementioned goals, which we hope to address in the future.
Notably, the use of a $\Delta$-learning approach currently requires two MLP models. 
While we have been able to prevent the expected doubling of energy and force inference costs, it still (1) increases the labor required to develop two MLPs simultaneously and (2) there may be an accumulation of errors from summing two models which we have not investigated in detail.
Moreover, the lack of long-range electrostatics limit these models' application to interfacial systems~\cite{fitznerManyFacesHeterogeneous2015,finneyMultiplePathwaysNaCl2022,oneillCrumblingCrystalsDissolution2024} -- where the broken translational symmetry is not well described by purely short-ranged models \cite{niblettLearningIntermolecularForces2021, yueWhenShortrangeAtomistic2021}.
This (literal) shortcoming also means that they cannot simultaneously describe both clusters and the bulk, as is possible with MBE-based models.
This $\Delta$-learning approach also implicitly assumes that the DFT baseline \textit{can} describe the long-range contributions that are missing from training purely on gas phase clusters.
Learning on isolated gas phase clusters for highly charged systems may be challenging to converge with DFT methods, which would ideally require some form of embedding procedure \cite{huangQuantumMechanicalEmbedding2011}.
Solving these two problems may require higher levels of theory as a baseline such as MP2 or RPA which are much closer to the CCSD(T) long-range behavior.

With the developments outlined within this work (along with their potential limitations described above in mind), we and others are primed to tackle more challenging problems beyond the structural and dynamical properties of liquid water.
We expect that it can be readily extended towards more complex aqueous solutions consisting of ions (\textit{i.e.}, electrolytes) \cite{oneillPairNotPair2024}.
We are currently applying this workflow to the ion pair association free energy of \ch{CaCO_3} in water, where a $\Delta$-MLP has been trained to clusters centered on the ions, with very good initial results that will be reported in forthcoming work.
Taking a further step in accurately and routinely predicting reactions occurring in various solvents would be very valuable, where high accuracy on the electronic structure front coupled with extensive sampling is necessary to reliably predict the possible reaction pathways \cite{a.youngTransferableActivelearningStrategy2021}.
Finally, while this work has only focused on a limited subset of liquid water properties, there is scope towards developing a more complete water model that can tackle other challenging properties.
These include the solid-liquid and liquid-air interfaces as well as describing the reactive nature of water and capturing response properties - the latter having seen recent promising progress \cite{jindalComputingBulkPhase2025}.

To summarize, we have proposed improvements to previous $\Delta$-learning approaches to reach CCSD(T)-quality MLPs for condensed phase simulations.
Here we summarize a blueprint as a starting point for developing and validating a $\Delta$-learned CCSD(T) quality MLP: 
\begin{enumerate}
\item Establish a robust NPT-capable baseline model through thorough sampling. For water this involved sampling configuration space across a range of densities and pressures, for both classical and quantum nuclei.
\item Generate cluster datasets to train a $\Delta$-MLP carved from bulk configurations by extracting progressively larger clusters, while retaining the smaller ones.
\item Converge observables with cluster size to identify a transferable cutoff. We find that $5.5\,$\AA{} clusters are sufficient for liquid water, which can be used as a starting point for other systems.
\item Use reliable basis sets and local approximations as practical defaults. For liquid water the jul-cc-pVQZ basis set along with normal/default thresholds for the LNO approximation is sufficient, serving as starting point for future work.
\item Use efficient architectures for the $\Delta$ correction to improve simulation performance.
\end{enumerate}

Using available structural (RDF) and transport (diffusion) data of liquid water, we have validated this approach, achieving good agreement with experiments.
Moreover, we show that the CCSD(T) MLP can handle constant pressure simulations, allowing for the density isobar of water -- a challenging property for DFT -- to be resolved.
We show our approach -- based on the progressive inclusion of water clusters of increasing size -- provides a systematic and straightforward means to converge and validate the dataset used to generate the the $\Delta$-MLP.
This strategy enables highly data-efficient training, requiring significantly fewer large (and computationally expensive) clusters than previous methods.
Finally, we perform benchmarks on the convergence on electronic structure parameters with basis set size and local approximation thresholds, as well as MLP architectures, proposing an efficient and cost-effective combination.
Together, these developments enable the efficient training of CCSD(T) MLPs, marking a further stepping stone towards routine condensed phase simulations at CCSD(T) accuracy.

\section{Methods}

\subsection{Machine learning potentials}

The machine learning potentials (MLP) in this work make use of the MACE architecture \cite{NEURIPS2022_4a36c3c5}, which achieves state-of-the-art accuracy and data-efficiency~\cite{kovacsEvaluationMACEForce2023} by employing equivariant message passing with local body-order descriptions of each atom.
We use a different choice of hyperparameters for the baseline and $\Delta$-MLPs.
The baseline DFT MLPs comprise 2 message passing layers with 128 channels and a $6\,$\AA{} cutoff, while this is reduced to 64 channels and a $4\,$\AA{} cutoff for the $\Delta$-MLP.
We show in Section~\ref{si-sec:architecture} of the SI and the Discussion that these reduced settings significantly decrease the inference cost of the $\Delta$-MLP with no compromise in accuracy.
For each dataset, 10\% was held back as a validation set.

\subsection{Density functional theory}
The DFT calculations required for both the baseline MLP and the $\Delta$-MLP was performed with FHI-AIMs~\cite{blumInitioMolecularSimulations2009}.
It can perform calculations under both periodic and open boundary conditions, allowing consistent treatment of the electronic structure of both the periodic and cluster calculations in the baseline and $\Delta$-MLP, respectively.
We used the \texttt{tight} family of basis sets and the (in-built) Grimme's D3~\cite{grimmeConsistentAccurateInitio2010} dispersion program in FHI-AIMs.
The periodic unit cell calculations were performed using the $\Gamma$-point.
As described in Section~\ref{si-sec:ml} of the SI, the dataset used to generate our periodic and $\Delta$-MLP models in Section \ref{si-sec:ml} of the SI, consisting of ${\sim}$1000 structures in the former and ${\sim}$7000 structures in the latter, all sampled from data generated at a wide range of pressures (with nuclear quantum effects).
This dataset was computed with revPBE-D3~\cite{zhangCommentGeneralizedGradient1998} (with zero damping), PBE-D3~\cite{perdewGeneralizedGradientApproximation1996} (with zero damping) and r$^2$SCAN~\cite{furnessAccurateNumericallyEfficient2020} as the baseline.

\subsection{Coupled cluster theory}
The dataset of CCSD(T) energies were calculated in both ORCA~\cite{neeseORCAQuantumChemistry2020} and MRCC~\cite{kallayMRCCProgramSystem2020}, utilizing the domain based local pair natural orbital (DLPNO)~\cite{riplingerEfficientLinearScaling2013,riplingerNaturalTripleExcitations2013,riplingerSparseMapsSystematic2016} and local natural orbital (LNO) approximations~\cite{nagyOptimizationLinearScalingLocal2018,gyevi-nagyIntegralDirectParallelImplementation2020}, respectively.
We have predominantly used the ORCA code, using conservative electronic structure parameters, namely ``TightPNO'' DLPNO thresholds, with resolution-of-identity approximations disabled for the initial Hartree-Fock (HF) calculations.
With the MRCC code, we have aimed to go towards more cost efficient electronic structure parameters, using the default ``normal'' LNO thresholds and enabling density-fitting to speed up the HF calculations.
For both codes, we used the jul-cc-pV$X$Z basis sets~\cite{papajakPerspectivesBasisSets2011}, consisting of the aug-cc-pV$X$Z basis set on the O and cc-pV$X$Z basis set on the H atoms, together with their corresponding density-fitting basis sets in the local CCSD(T) correlation treatment.
We performed complete basis set (CBS) extrapolations for the triple (TZ) and quadruple-zeta (QZ) basis sets, using parameters taken from Neese and Valeev~\cite{neeseRevisitingAtomicNatural2011}.

\subsection{Molecular dynamics}

With the resulting CCSD(T) MLP, molecular dynamics simulations were performed to obtain the following observables: density isobar (and the density at $298\,$K and $1\,$bar pressure), radial distribution function (RDF), and diffusion coefficient.
All classical simulations were performed using the Large-scale Atomic/Molecular Massively Parallel Simulator (LAMMPS)~\cite{thompsonLAMMPSFlexibleSimulation2022} code, coupled with the \texttt{Symmetrix} library \cite{wittSymmetrixAvailableHttps2025}, using the \texttt{symmetrix/mace} pair style in tandem with the \texttt{hybrid/overlay} pairstyle to sum the periodic and $\Delta$-MLPs.
All simulation boxes contained 126 waters and were performed at 298 K.
Classical simulations used a 0.5 fs timestep.
The density was obtained from simulations in the isothermal isobaric ensemble (NPT) at a pressure of 1 bar, with a barostat relaxation time of 1 ps.
All density simulations were run for at least 500 ps, with block averaging to obtain the error bar.
Radial distribution functions and diffusion coefficients were obtained from simulations in the canonical (NVT) ensemble.
For all convergence tests, these were computed at the experimental density, with a box size of 15.577 \AA{} to ensure consistency.
For the final model production simulations, the RDF and diffusion coefficient were obtained from simulations at the computed density from the NPT simulations.
In all cases, the CSVR thermostat \cite{bussiCanonicalSamplingVelocity2007} was used, with a temperature of 298 K and a temperature relaxation time of 0.1 ps.
Path integral molecular dynamics simulations were performed using the i-PI code\cite{litmanIPI30Flexible2024} interfaced with LAMMPS \cite{thompsonLAMMPSFlexibleSimulation2022} and symmetrix \cite{kovacsMACEOFFShortRangeTransferable2025}.
Densities were obtained from ring polymer molecular dynamics (RPMD)~\cite{craigQuantumStatisticsClassical2004} simulations in the isobaric isothermal ensemble using 32 beads, with a 0.25 fs timestep, with each replica at least 200 ps long.
RDFs and difusion coefficients were computed in the canonical ensemble using thermostatted RPMD (T-RPMD)~\cite{rossiHowRemoveSpurious2014} at 298 K with a 0.25 fs timestep, using the computed density.
An average of 9 trajectories of 100 ps each was taken for the final diffusion coefficient, obtained from the mean squared displacement of the centroid of the ring polymer.

\section*{Data availability}
All data required to reproduce the findings of this study will be made available upon publication of this study.

\section*{Code availability}
All simulations were performed with publicly available simulation software (\texttt{ACEsuit}, \texttt{LAMMPS}, \texttt{Symmetrix}).

\section*{Acknowledgments}
N.O.N acknowledges financial support from the Gates Cambridge Trust and is grateful for an International Exchange Grant funded by the Royal Society. The Flatiron Institute is a division of the Simons Foundation. W.C.W. acknowledges support from the EPSRC (Grant EP/V062654/1). A.M. acknowledges support from the European Union under the ``n-AQUA'' European Research Council project (Grant No. 101071937). J.D.G thanks the Australian Research Council for funding under grant FL180100087, as well as the Pawsey Supercomputing Centre and National Computational Infrastructure for computing resources. C.S. acknowledges financial support from the Royal Society, grant number RGS/R2/242614. We are grateful for computational support and resources from the UK national high-performance computing service, Advanced Research Computing High End Resource (ARCHER2). Access for ARCHER2 were obtained via the UK Car-Parrinello consortium, funded by EPSRC grant reference EP/P022561/1. We also acknowledge the EuroHPC Joint Undertaking for awarding this project access to the EuroHPC supercomputer LEONARDO, hosted by CINECA (Italy) and the LEONARDO consortium through an EuroHPC Regular Access call. This work was also performed using resources provided by the Cambridge Service for Data Driven Discovery (CSD3) operated by the University of Cambridge Research Computing Service (www.csd3.cam.ac.uk), provided by Dell EMC and Intel using Tier-2 funding from the Engineering and Physical Sciences Research Council (capital grant EP/T022159/1), and DiRAC funding from the Science and Technology Facilities Council (www.dirac.ac.uk)

\end{document}

\title{\mytitle}

\date{\today}

\author{Niamh O'Neill}%
\email{nco24@cam.ac.uk}
\affiliation{%
Yusuf Hamied Department of Chemistry, University of Cambridge, Lensfield Road, Cambridge, CB2 1EW, UK
}
\affiliation{%
Cavendish Laboratory, Department of Physics, University of Cambridge, Cambridge, CB3 0HE, UK
}
\affiliation{%
Lennard-Jones Centre, University of Cambridge, Trinity Ln, Cambridge, CB2 1TN, UK
}
\author{Benjamin X. Shi}%
\email{mail@benjaminshi.com}
\affiliation{Initiative for Computational Catalysis, Flatiron Institute, 160 5th Avenue, New York, NY 10010}

\author{William Baldwin}%
\affiliation{%
Lennard-Jones Centre, University of Cambridge, Trinity Ln, Cambridge, CB2 1TN, UK
}
\affiliation{%
Department of Engineering, University of Cambridge, Cambridge, CB3 0HE, UK
}

\author{William C. Witt}%
\affiliation{%
Harvard John A. Paulson School of Engineering and Applied Sciences, Harvard University, Cambridge, MA, USA
}
\author{G\'abor Cs\'anyi}
\affiliation{%
Lennard-Jones Centre, University of Cambridge, Trinity Ln, Cambridge, CB2 1TN, UK
}
\affiliation{%
Department of Engineering, University of Cambridge, Cambridge, CB3 0HE, UK
}
\author{Julian D. Gale}%
\affiliation{%
School of Molecular and Life Sciences, Curtin
University, PO Box U1987, Perth, Western Australia 6845, Australia
}
\author{Angelos Michaelides}%
\affiliation{%
Yusuf Hamied Department of Chemistry, University of Cambridge, Lensfield Road, Cambridge, CB2 1EW, UK
}
\affiliation{%
Lennard-Jones Centre, University of Cambridge, Trinity Ln, Cambridge, CB2 1TN, UK
}
\author{Christoph Schran}%
\email{cs2121@cam.ac.uk}
\affiliation{%
Cavendish Laboratory, Department of Physics, University of Cambridge, Cambridge, CB3 0HE, UK
}
\affiliation{%
Lennard-Jones Centre, University of Cambridge, Trinity Ln, Cambridge, CB2 1TN, UK
}

\maketitle
\tableofcontents

\newpage
Many of the SI tests are performed in the same self-consistent spirit as the convergence tests in the main text for computational efficiency, therefore for a given property, the final 'converged' result may not be the basis set converged CCSD(T) result. For clarity, the level of theory (CCSD(T) basis set/ local approximation thresholds) is given in the caption of each Figure. Furthermore, all tests on the RDF in the SI are done at the experimental density (0.997 g/cm$^3$ at 298 K) to enable consistent comparisons.
\section{Machine learning}\label{sec:ml}
\subsection{Dataset}
The $\Delta$-learning approach used in this work requires 2 machine learning potentials (MLPs) and therefore two datasets.
The baseline MLP was trained on periodic boxes containing 126 waters, labeled with DFT forces and energies.
The $\Delta$-MLP was trained on gas phase clusters cut from periodic configurations and labeled with the energy difference between CCSD(T) and DFT.
The final CCSD(T) MLP is given as a sum of the periodic baseline model plus the $\Delta$-MLP.
Note that any time we refer to the CCSD(T) MLP, we are referring to this total sum of the baseline plus $\Delta$-MLPs.
In practice, configurations for both model datasets were sampled from the same simulations and the datasets was generated over multiple generations. 

An additional challenge comes from the extent of the differences between the DFT baseline and CCSD(T).
For example, revPBE-D3 predicts a density roughly 10 \% smaller than experiment \cite{monterodehijesDensityIsobarWater2024} (the target for the CCSD(T) model), and therefore care should be taken to ensure sufficiently diverse density sampling to obtain stable reliable models, as well as sampling from both DFT and CCSD(T) configurations.

In summary, the following simulations were sampled, with each used to provide reference configurations for both baseline and $\Delta$-MLPs:
NPT pressure scan at DFT level spanning both positive and negative pressures (specifically -1500, -500, -300, 1, 2, 500, 1000, 4000, 8000 bar).
This gave an initial CCSD(T) MLP, which was then used to sample again NPT with the same positive and negative pressures, this time ensuring CCSD(T) level water structures were covered in the dataset.
Finally to ensure the model could robustly describe NQEs, a final sampling of PIMD NVT simulations over a density scan (0.85, 0.9, 0.95, 1.0, 1.05 g/cm$^{3}$) was done to give the final CCSD(T) model.

\subsection{$\Delta$-MLP}
One challenge with the $\Delta$-learning strategy is the significantly reduced information content training on energies alone, with each configuration containing only a single energy.
This is in contrast to the standard approach for fitting DFT-level MLPs, where energy gradients (forces) provide significant additional information on the slope of the potential energy surface. 
Therefore, we have specifically explored the number of clusters required to reliably predict bulk properties from gas phase clusters (as well as the size of clusters as discussed in the main text) and this is discussed more in Section \ref{sec:data_efficiency}.

The clusters were cleaved out from the same set of structures used for the periodic DFT dataset.
An O atom was randomly selected within each structure (or its supercell depending on the cluster radius $r_c$), and all O atoms within $r_c$ around this O atom were selected.
The cluster was then formed by then incorporating all H atoms bonded to the selected O atoms.

\subsection{Validation}
We have performed a series of validation tests on the models to test the completeness of the dataset, MLP architectures and seed and baseline dependencies.
As in the main, we directly benchmark against condensed phase properties. For the majority of these tests, we show results for the density and radial distribution function (RDF), since reliably computing the diffusion coefficient requires additional computational effort, which would be significant for the roughly one hundred models we have trained for validation purposes.
Nevertheless, the extent of structuring of the RDF has been shown to be a good proxy for the water self-diffusion coefficient \cite{chenDataEfficientMachineLearning2023}, and therefore the properties we validate against still represent a highly thorough suite of tests. Moreover, testing directly on condensed phase observables makes a much more direct connection than the typical benchmarks performed on gas phase properties of small clusters.

\subsubsection{\label{sec:seed_dep}Seed Dependence}
\begin{figure}[h]
    \centering
    \includegraphics[width=0.67\linewidth]{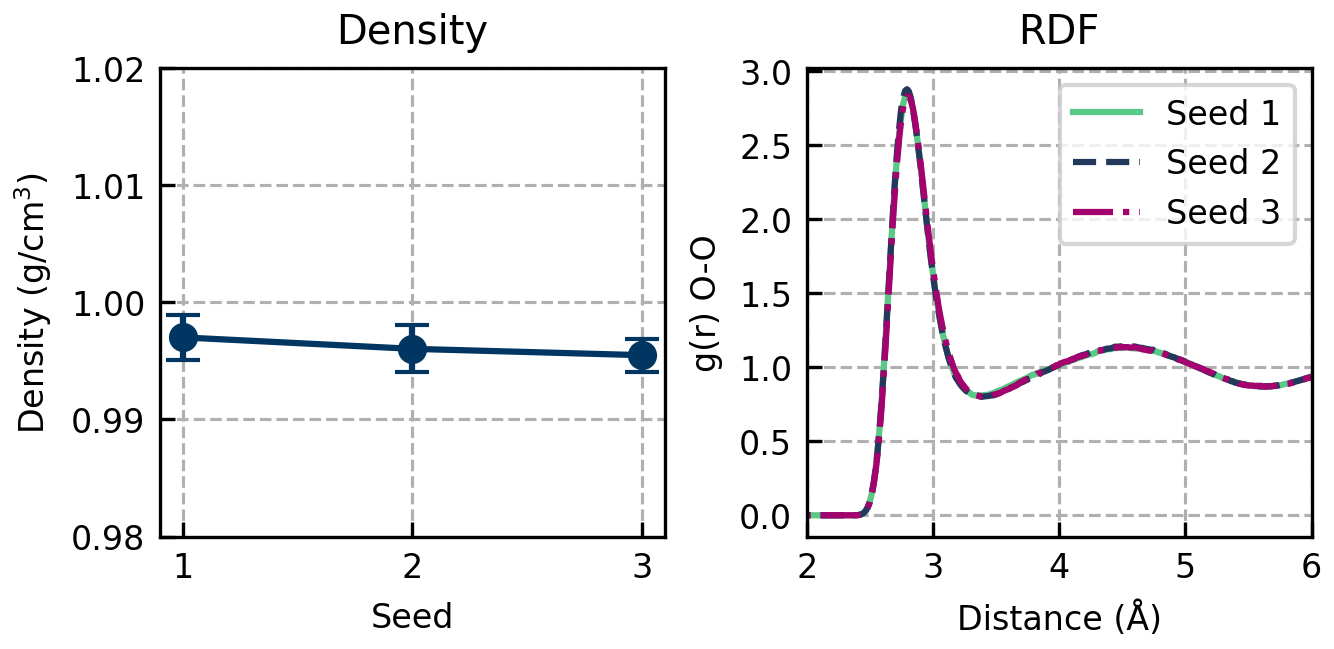}
    \caption{\textbf{Seed dependence.} Comparison of three different seeds for the $\Delta$ model on the CCSD(T) prediction of density and RDF. [Level of theory of CCSD(T): QZ/TightPNO]}
    \label{fig:seed}
\end{figure}

To test the completeness of the dataset, we trained several $\Delta$-MLP models (with the r$^2$SCAN baseline) with different train-test-validation splits as well as seeds.
The resulting thermodynamic properties of the different seeds is shown in Figure~\ref{fig:seed}.
We find that all seeds correct the shortcomings of r$^2$SCAN with respect to experiment, with minor variations within 0.03 g/cm$^3$ on the density. 

\subsection{\label{sec:architecture}Architecture}
We use the MACE framework \cite{NEURIPS2022_4a36c3c5} to train MLPs for both the baseline and $\Delta$-MLPs.
Choosing hyperparameters for the baseline MLP are now routine and straightforward (and are detailed in the main Methods section) and the focus in this section are to find optimal hyperparameters for the $\Delta$-MLP.
In particular, previous work has highlighted that a shorter range cutoff may be needed/sufficient to fit this potential \cite{meszarosShortRangeDMachineLearning2025a}, which can help to lower the overall computational cost.
Therefore the most prominent hyperparameter to consider is the cutoff, but we also consider the inclusion of 2 message-passing layers and the number of channels.
The messages can also be chosen to be equivariant or invariant (L=0 or L=1).

\subsubsection{Number of message passing layers and number of channels}
\begin{figure}[h!]
    \centering
    \includegraphics[width=1.0\linewidth]{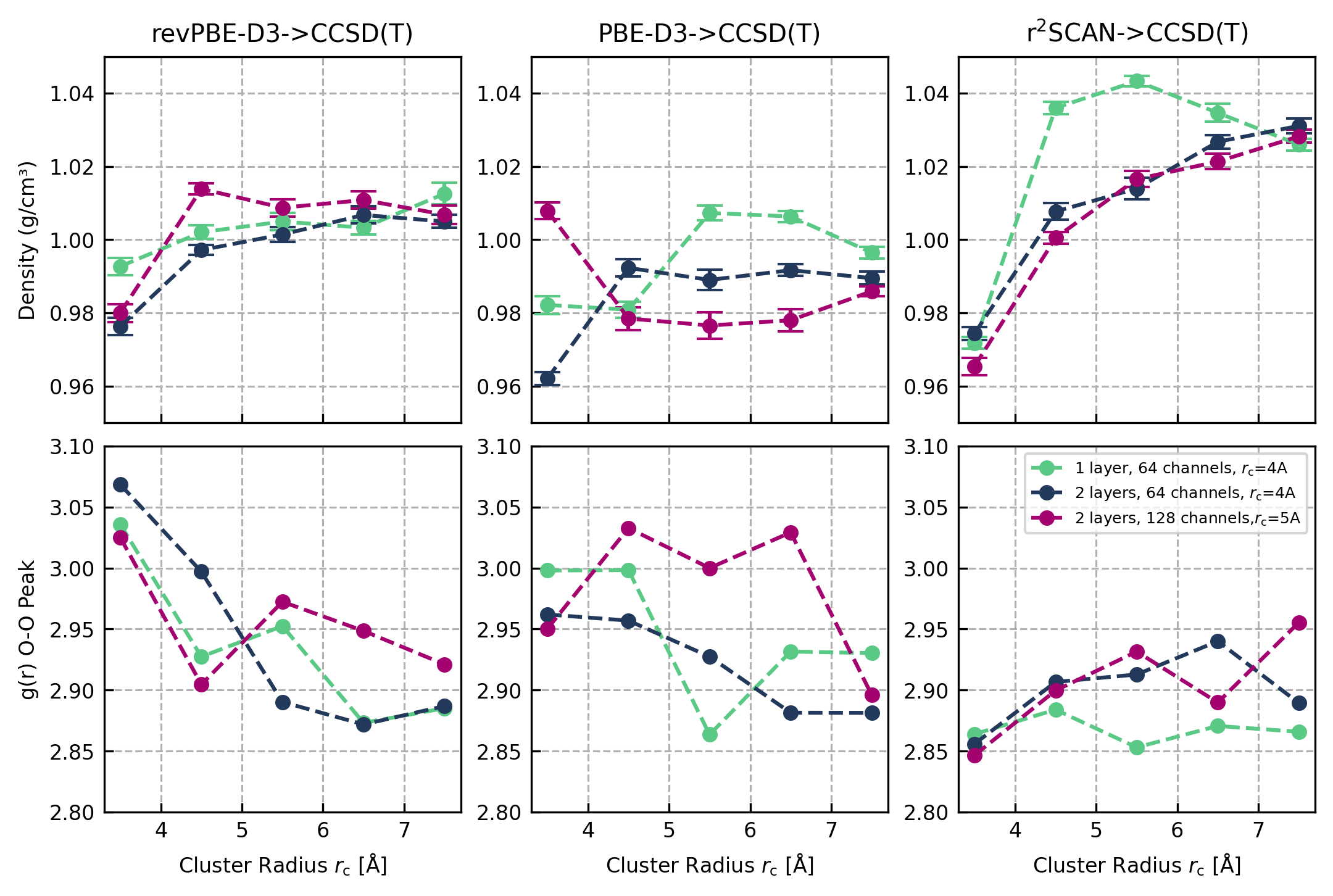}
    \caption{\textbf{Architecture and dataset relationship.} Comparison of density and RDF peak position for the three different DFT baselines for various MACE architectures for the $\Delta$-MLP as a function of the largest size of clusters in the dataset. Both 1 and 2-layer models are shown, where we have also varied the number of channels and cutoff radius (r$_c$) for the 2-layer model. [Level of theory of CCSD(T): CBS(DZ/TZ) NormalLNO]}
    \label{fig:architecture}
\end{figure}

We first explore the effect of the number of message passing layers (1 or 2), with the number of channels (64 vs 128) and cutoff (4 vs 5$\,$\AA).
To get a further handle on the variation, we perform this test using the three different baselines.
We test the effect of these architectures on the density and RDF as a function of the (cumulative) cluster sizes in the training set as shown in Figure \ref{fig:architecture}.
In the limit of large clusters (7.5$\,$\AA), the predicted densities for all models agrees within error bars.
However, for smaller cluster sizes at the sizes we are targeting for the final model (5.5$\,$\AA{} and smaller), the single layer model shows larger fluctuations for the r$^2$SCAN and PBE-D3 baselines, and so we chose the 2-layer, 64 channel model with a $4\,$\AA{} cutoff for the final architecture.

\subsubsection{Cutoff}
\begin{figure}[h!]
    \centering
    \includegraphics[width=0.67\linewidth]{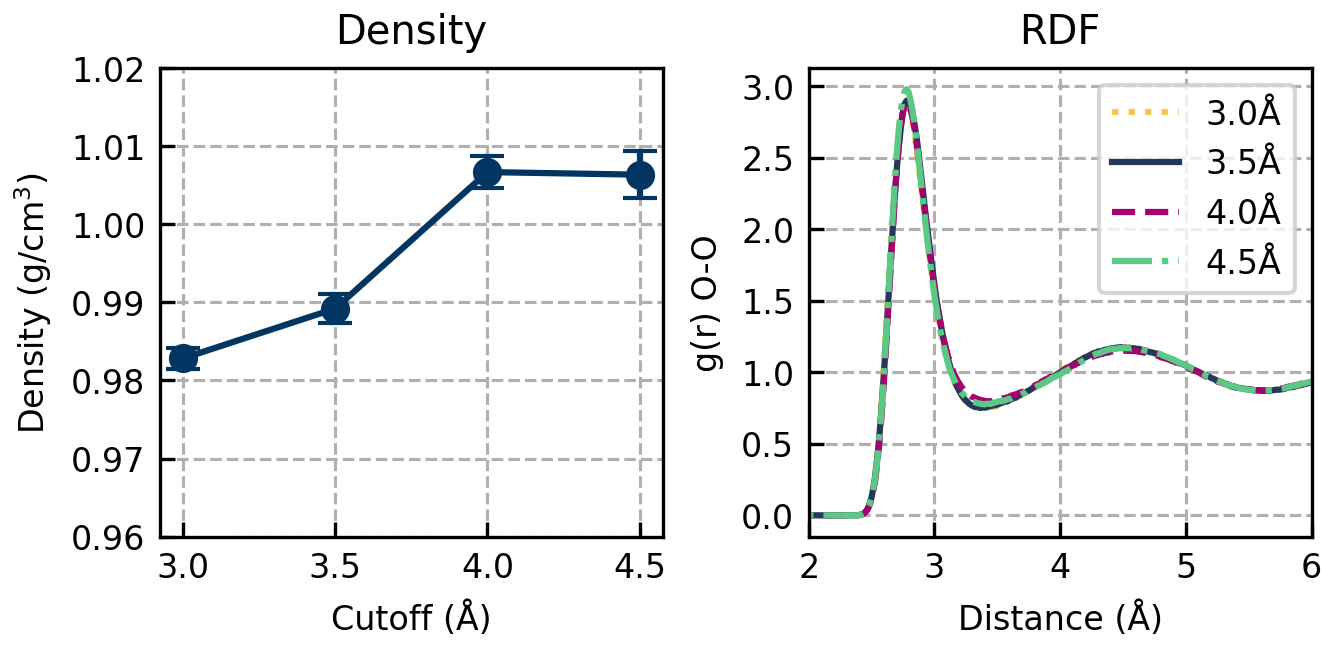}
    \caption{\textbf{Cutoff of $\Delta$-MLP.} Effect of cutoff for 2 layer $\Delta$ model with 5.5$\,$\AA{} the maximum cluster radius in the dataset using revPBE-D3 as a baseline. [Level of theory of CCSD(T): QZ/TightPNO ]}
    \label{fig:cutoff}
\end{figure}

We show in Figure 2 in the main text, that the dataset containing up to 5.5$\,$\AA{} clusters reliably reproduces the limit of larger cluster sizes.
For this 5.5$\,$\AA{} cluster dataset, we have also tested the effect of the MACE cutoff for the 2 layer, 64 channel model selected from previous section.
We test cutoffs from 3.0$\,$\AA{} to 4.5$\,$\AA{} in steps of 0.5$\,$\AA{}, such that the complete receptive field of the first layer is filled.
Figure \ref{fig:cutoff} shows that while the RDF is already well described by the smallest receptive field, the density requires a larger cutoff of 4.0$\,$\AA{}.

\subsubsection{L0 vs L1}

\begin{figure}[h]
    \centering
    \includegraphics[width=0.67\linewidth]{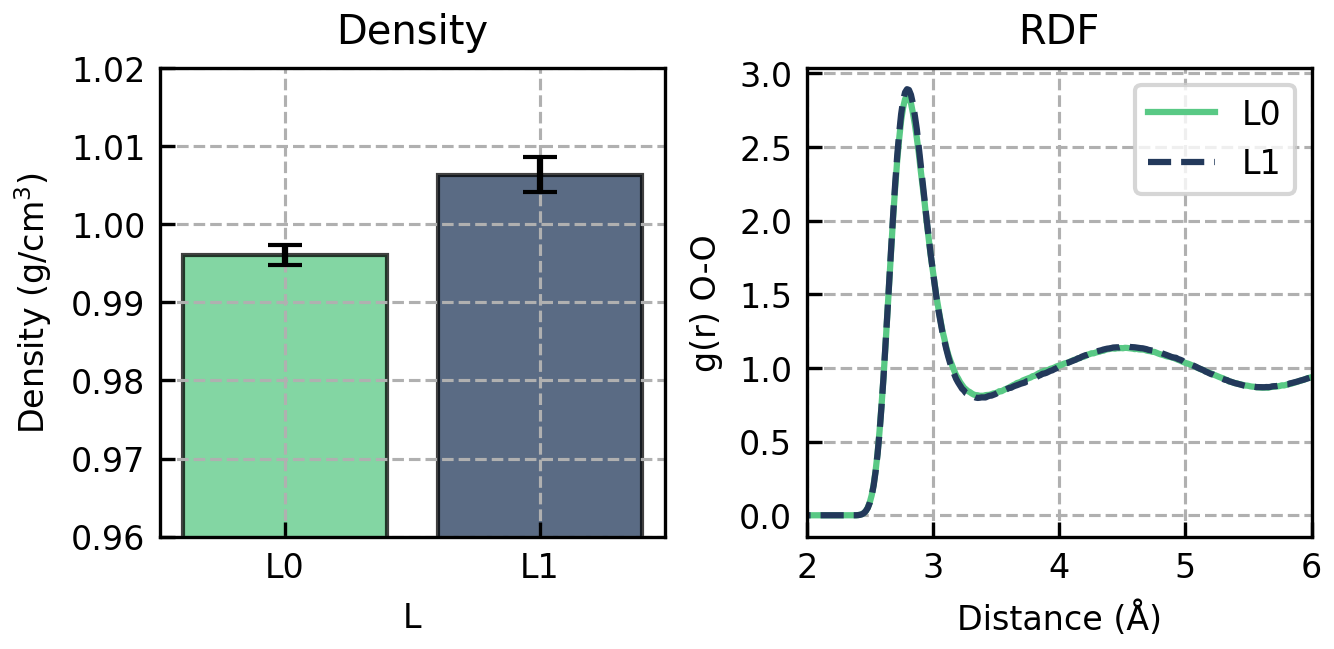}
    \caption{\textbf{Comparison of MACE message-passing type.} Effect of invariant (L=0) vs equivariant (L=1) messages for $\Delta$-MLP on CCSD(T) density and RDF using revPBE-D3 as a baseline. [Level of theory of CCSD(T): QZ/TightPNO]}
    \label{fig:l0l1}
\end{figure}

In Figure~\ref{fig:l0l1}, we compare the effect of utilizing either invariant (L=0) or equivariant (L=1) messages for a MACE cutoff of 4.0$\,$\AA{} and 64 channels. We find that differences are minor for both density and RDF, highlighting that invariant messages are sufficient with our dataset.

\section{Overall validation}
\subsection{DFT to DFT validation: revPBE-D3 to r2SCAN; revPBE-D3 to PBE-D3}\label{sec:dft2dft}
\begin{figure}[h]
    \centering
    \includegraphics[width=1.0\linewidth]{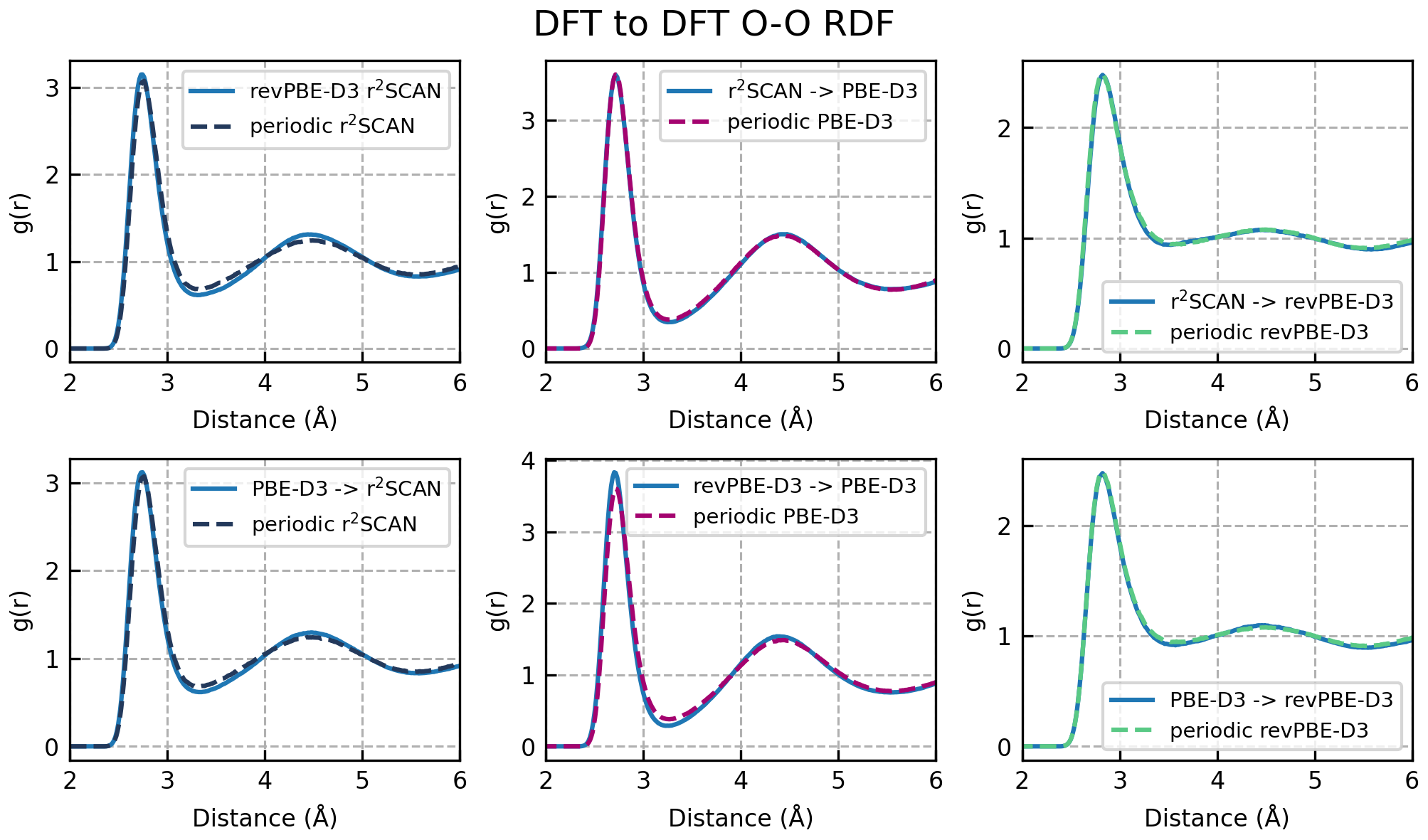}
    \caption{\textbf{DFT to DFT validation on the O-O RDF} O-O RDFs for all permutations going between r$^2$SCAN, revPBE-D3 and PBE-D3. Dashed lines give the periodic model result, and solid line is the delta model with $A \rightarrow B$
 corresponding to $\mathrm{baseline} \rightarrow \Delta \, \mathrm{model}$
.}
    \label{fig:OO-DFT}
\end{figure}

\begin{figure}
    \centering
    \includegraphics[width=1.0\linewidth]{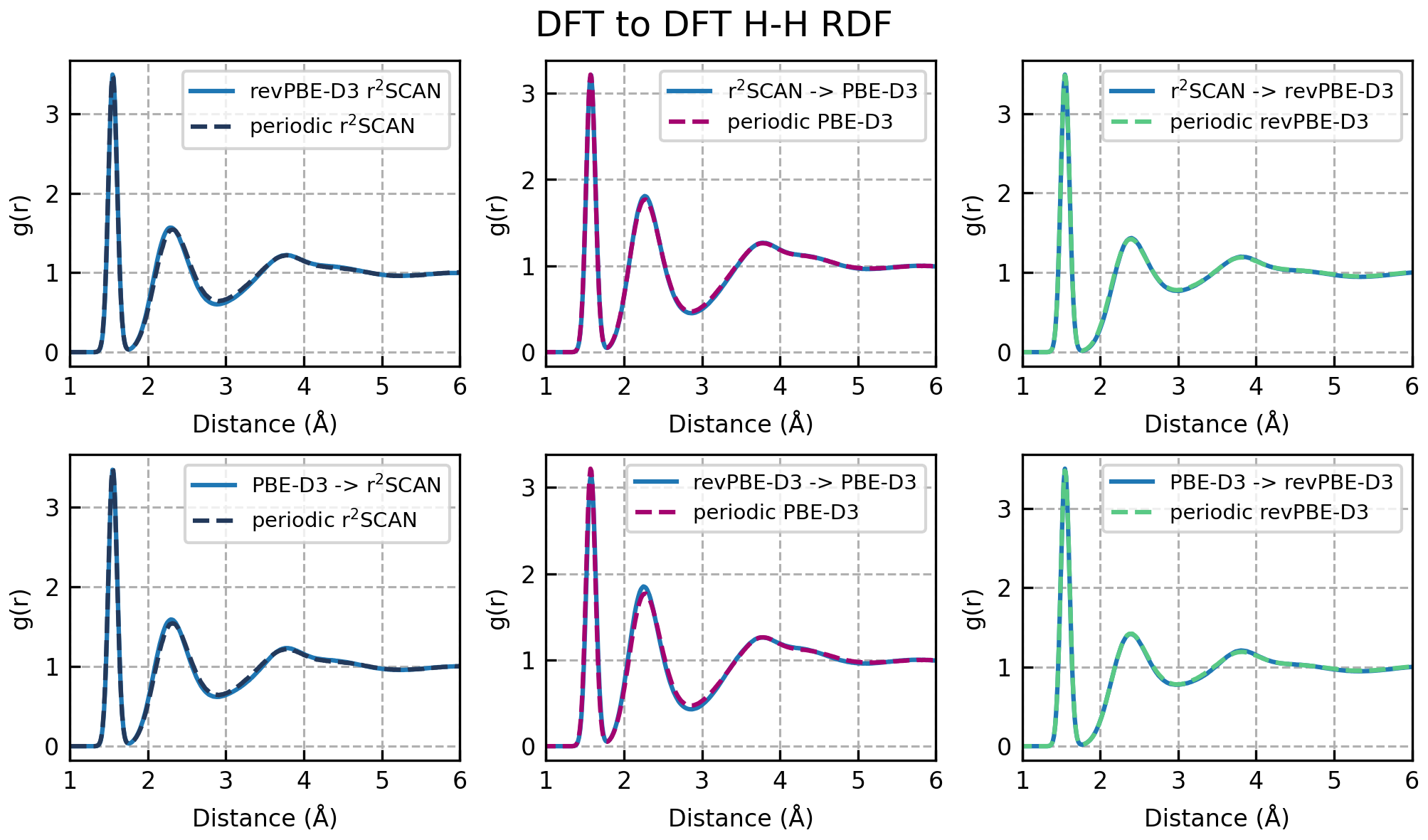}
    \caption{\textbf{DFT to DFT validation on the H-H RDF.} H-H RDFs for all permutations going between r$^2$SCAN, revPBE-D3 and PBE-D3.}
    \label{fig:HH-DFT}
\end{figure}

\begin{figure}
    \centering
    \includegraphics[width=1.0\linewidth]{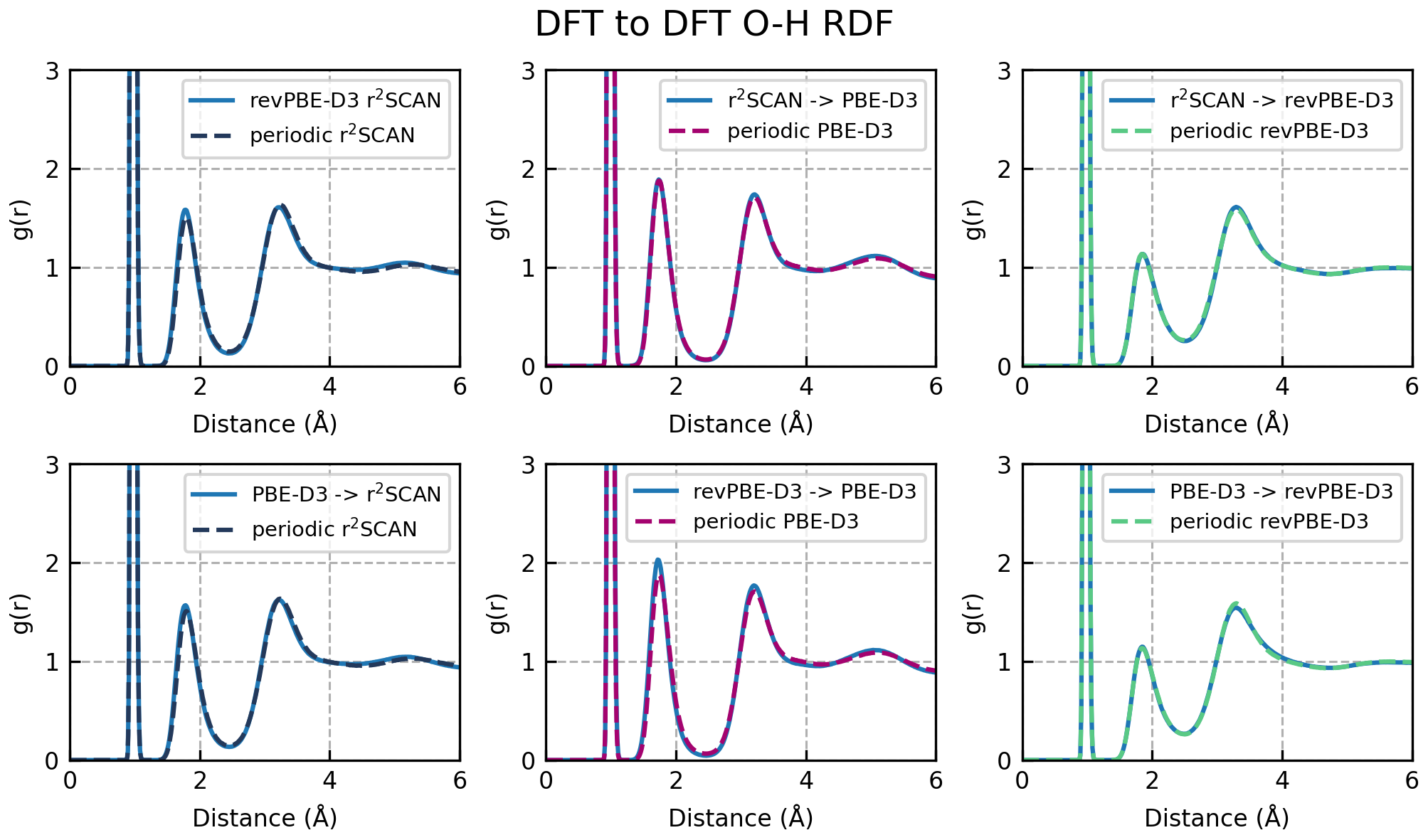}
    \caption{\textbf{DFT to DFT validation on the O-H RDF.} O-H RDFs for all permutations going between r$^2$SCAN, revPBE-D3 and PBE-D3.}
    \label{fig:OH-DFT}
\end{figure}

As an additional validation to the self-consistant convergence procedure described in the main text, here we validate the $\Delta$-learning approach by showing that we can learn between different levels of DFT.
In this validation, we can directly compare to the target periodic result, obtained from MLPs trained on periodic data.
We have chosen three DFT functionals, which give varying predictions of physical properties as well as the different physical attributes (\textit{i.e.}, the incorporation of dispersion corrections).
Figures \ref{fig:OO-DFT}, \ref{fig:HH-DFT} and \ref{fig:OH-DFT} all show the $\Delta$ approach learns the difference between these different DFT levels for the O-O, H-H and O-H RDFs, reliably reproducing the reference periodic target.
We note that there are slight differences between the revPBE-D3 to PBE-D3 model compared to the periodic PBE-D3 result.
This may be a result of not having included PBE-D3 water structures in the dataset, as further discussed in Section \ref{sec:dft2ccsdt} below.

\subsection{DFT to CCSD(T) validation: revPBE-D3 to CCSD(T); r\textsuperscript{2}SCAN to CCSD(T); PBE-D3 to CCSD(T)}\label{sec:dft2ccsdt}

\begin{figure}[h]
    \centering
    \includegraphics[width=1.0\linewidth]{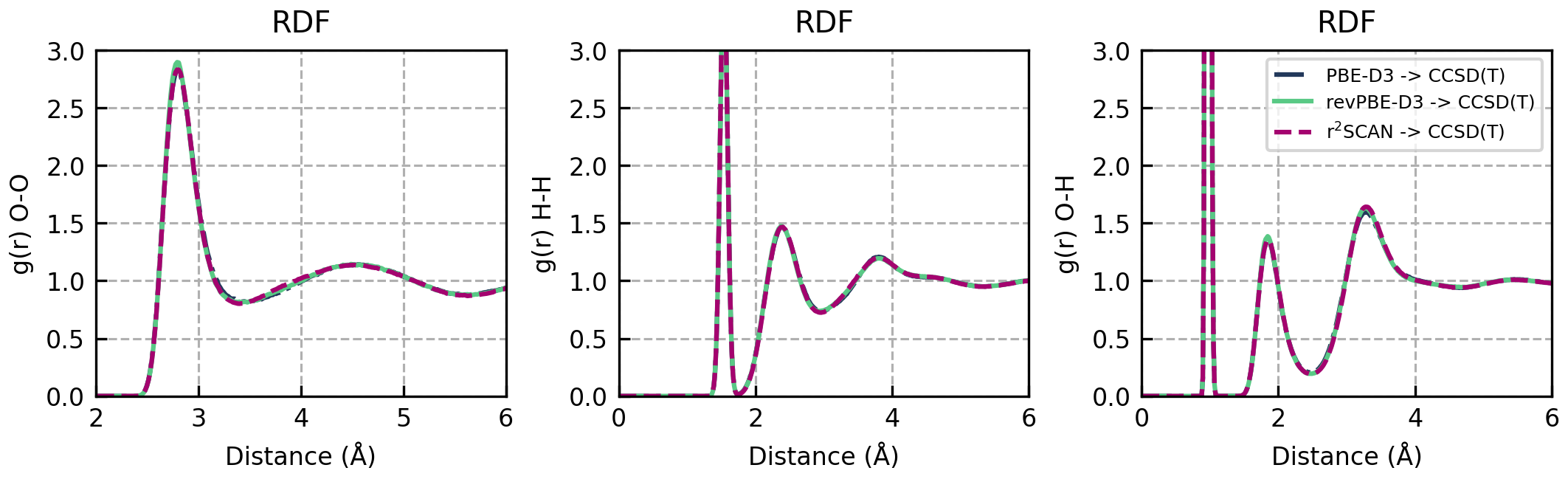}
    \caption{\textbf {DFT to CCSD(T) validation.} Predictions of CCSD(T) RDFs based on three DFT baselines (PBE-D3, r$^2$SCAN  and revPBE-D3). [Level of theory of CCSD(T): QZ/ TightPNO]}
    \label{fig:dft2ccsdt_rdf}
\end{figure}

\begin{figure}[h]
    \centering
    \includegraphics[width=0.33\linewidth]{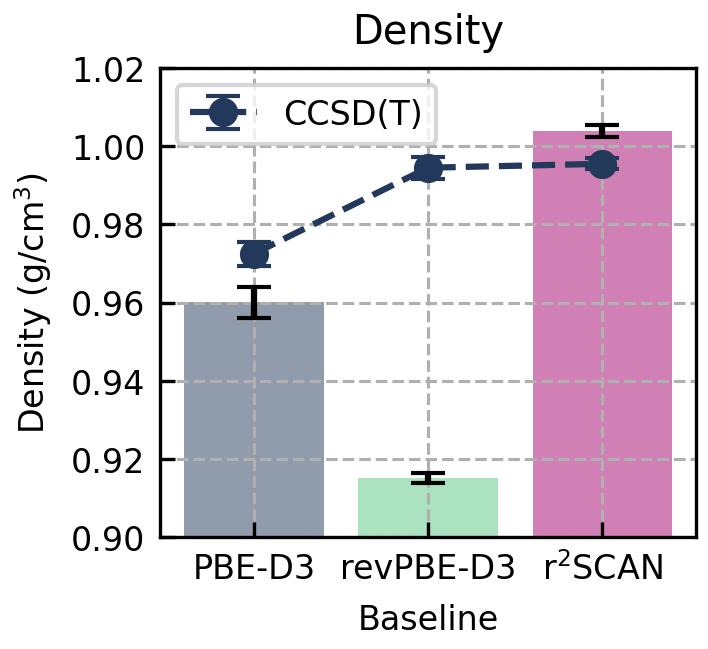}
    \caption{\textbf {DFT to CCSD(T) validation.} Predictions of CCSD(T) density (dark blue) based on three DFT baselines (PBE-D3, r$^2$SCAN  and revPBE-D3). The corresponding baseline DFT density is shown in the barplot.  [Level of theory of CCSD(T): QZ/ TightPNO]}
    \label{fig:dft2ccsdt_density}
\end{figure}

To further validate the reliability of the $\Delta$ learning approach to learn the CCSD(T) reference, Figures \ref{fig:dft2ccsdt_rdf} and \ref{fig:dft2ccsdt_density} shows the combined $\Delta$-MLP predictions with different DFT baselines (revPBE-D3, r$^2$SCAN and PBE-D3).
We show that both revPBE-D3 and r$^2$SCAN baselines are in excellent agreement for the RDFs and density, suggesting the robustness of our framework, since revPBE-D3 and r$^2$SCAN are very different, both in terms of functional design philosophy and also prediction of physical properties of liquid water.
We also show that the RDF is least sensitive to the chosen baseline, with all three baselines in excellent agreement.
The CCSD(T) MLP with PBE-D3 as a baseline underpredicts the density by ${\sim}$2 \%.
While this may indicate that PBE-D3 is inherently harder to learn from, it should also be noted that we did not explicitly include the DFT (PBE-D3) water structures in the $\Delta$-MLP dataset.
We have found that including explicitly both DFT and CCSD(T) water in both the periodic and $\Delta$ datasets resulted in more stable models.
Nevertheless, this PBE-D3 result indicates that reasonable results can still be obtained from less judicious sampling.
However, for the most quantitatively predictive models, care should be taken to ensure the datasets span both the baseline and target levels of theory.

\begin{figure}[h]
    \centering
    \includegraphics[width=0.5\linewidth]{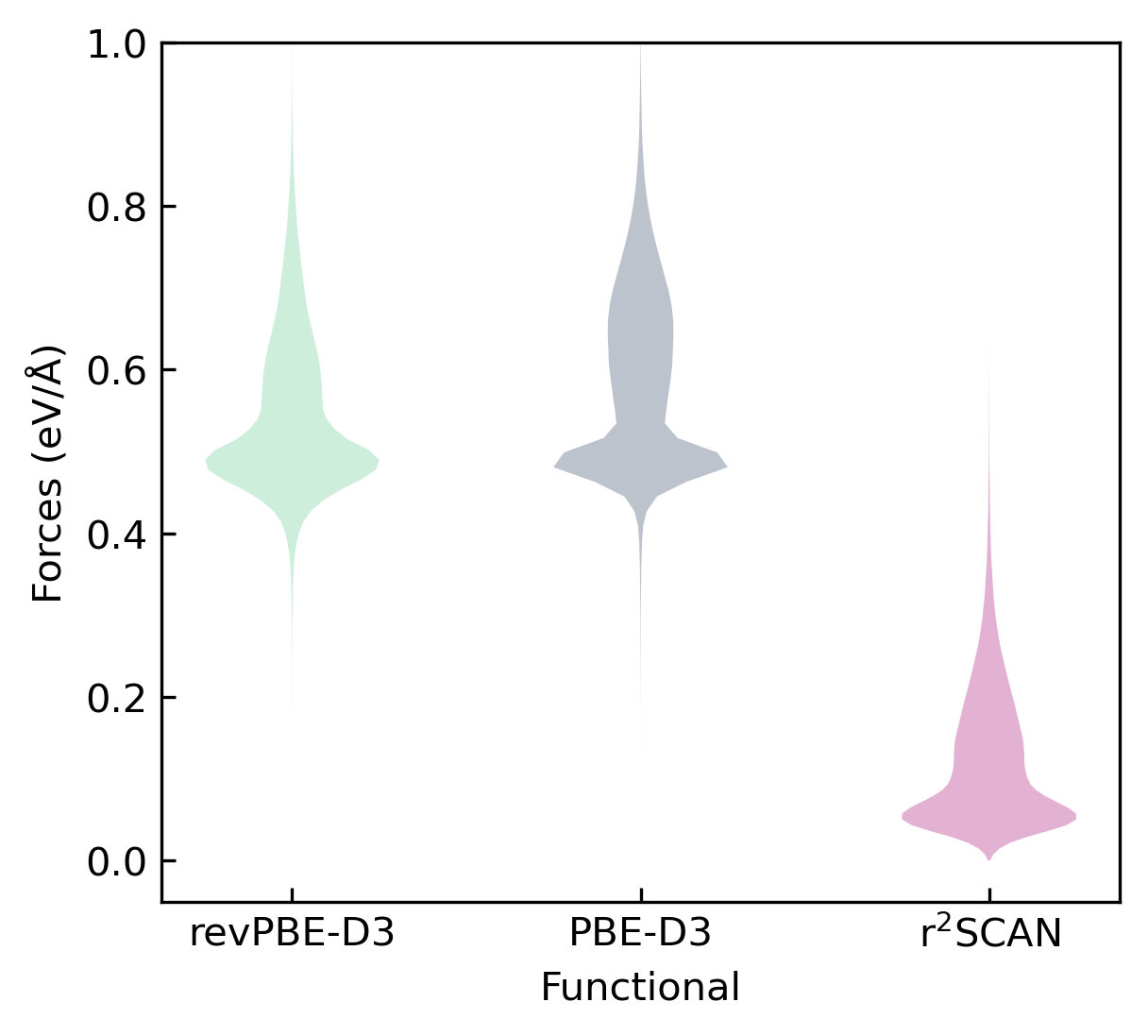}
    \caption{\textbf{Predicted force magnitude of the $\Delta$-MLP.} Violin plot of the magnitude of the atomic forces predicted by the $\Delta$-MLPs as a function of their baseline.}
    \label{fig:forces}
\end{figure}

In Figure~\ref{fig:forces}, we plot violin plots of the distribution of the predicted forces on the baseline dataset for the three $\Delta$-MLPs.
It can be seen that the $\Delta$-MLP with an r$^2$SCAN baseline predicts much smaller forces, indicating that r$^2$SCAN predicts closer energies and forces to CCSD(T) and so the $\Delta$-MLP has an easier learning task going between r$^2$SCAN to CCSD(T).

\section{Electronic Structure-computational details}

\subsection{CCSD(T)}
For completeness, here we repeat the CCSD(T) settings described in the Methods Section of the Main text.
The CCSD(T) calculations, using local approximations, were performed on both the MRCC~\cite{kallayMRCCProgramSystem2020} and ORCA~\cite{neeseORCAQuantumChemistry2020} program.
We used the domain-based local pair natural orbital (DLPNO) approximation~\cite{riplingerEfficientLinearScaling2013,riplingerNaturalTripleExcitations2013,riplingerSparseMapsSystematic2016} within ORCA for the final simulations shown in the main text (Main: Figures 3 and 4 and Table 1).
We used relatively conservative and accurate settings in ORCA, using the ``TightPNO'' DLPNO thresholds for the correlation energy calculations and turning off the RIJCOSX approximation for the Hartree-Fock calculations.
For many of the tests, particularly when demonstrating the cluster size convergence in Figure 2 of the main text, when validating the use of local approximations in Section~\ref{sec:mrcc_orca_valid}, and demonstrating computational cost savings from employing looser local approximation thresholds in Section~\ref{sec:data_efficiency}, we use the local natural orbital (LNO) approximation~\cite{nagyOptimizationLinearScalingLocal2018,nagyApproachingBasisSet2019} to CCSD(T) in MRCC.
In particular, we utilize density-fitting with the HF calculations as well as the standard ``normal'' LNO thresholds (which we have dubbed ``NormalLNO'' throughout this SI).

We have used the Dunning family~\cite{petersonAccurateCorrelationConsistent2002} of correlation consistent basis sets, where aug-cc-pV$X$Z was used on the O atom and cc-pV$X$Z was used on the H atom -- dubbed jul-cc-pV$X$Z --  with $X$ representing its size in terms of double (DZ), triple (TZ) or quadruple (QZ) zeta.
We consider the CBS(DZ/TZ) and CBS(TZ/QZ), which involve a two-point complete basis set (CBS) extrapolation, using parameters taken from Neese and Valeev~\cite{neeseRevisitingAtomicNatural2011}, for the enclosed pair of basis functions.
We use the def2-QZVPP-RI-JK auxiliary basis function for density-fitting/resolution-of-identity  Hartree--Fock (HF) computations in MRCC, and the resolution-of-identity auxiliary basis sets from Weigend~\cite{weigendRIMP2OptimizedAuxiliary1998,hellwegOptimizedAccurateAuxiliary2007} corresponding to the AO basis sets for subsequent local CCSD(T) calculations in ORCA and MRCC.

\subsection{\label{sec:mrcc_orca_valid}Validating local CCSD(T) approximations in MRCC and ORCA}
\begin{figure}[h!]
    \centering
    \includegraphics[width=0.67\linewidth]{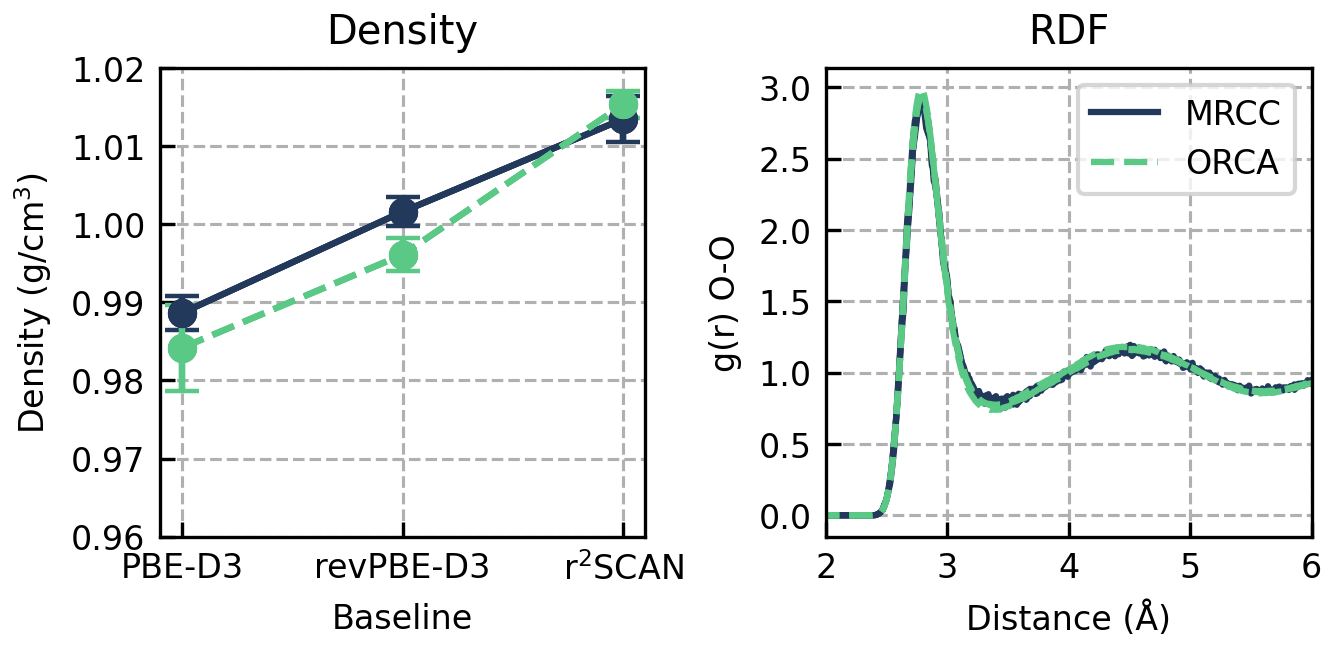}
    \caption{\textbf{Comparison between DLPNO and LNO CCSD(T).} Comparison of density and O-O RDF predictions for the CCSD(T) MLP using MRCC with \texttt{normal} LNO thresholds and density fitting in the HF for cbs(DZ/TZ) basis set for the $\Delta$-MLP.}
    \label{fig:dztz}
\end{figure}

As discussed in Section~\ref{sec:data_efficiency}, we have used relatively conservative DLPNO and HF settings for the ORCA calculations, while we considered cheaper settings, employing density-fitting in the HF calculations and only ``normal'' LNO thresholds, for the MRCC calculations.
As we show in Figure \ref{fig:dztz}, these less-conservative MRCC settings lead to observables which are in agreement with those from the conservative ORCA settings.
This agreement validates the both flavors of local approximations used in MRCC (LNO) and ORCA (DLPNO).
This sets the stage for using MRCC to study more complex systems, since its use of less conservative settings result in a computational cost is almost two orders of magnitude lower, as discussed in Section~\ref{sec:data_efficiency}.

\section{Molecular dynamics computational details}

\subsection{Simulation details for classical molecular dynamics}
All classical simulations were performed using the Large-scale Atomic/Molecular Massively Parallel Simulator (LAMMPS) code, coupled with the \texttt{Symmetrix} library \cite{wittSymmetrixAvailableHttps2025}, using the \texttt{symmetrix/mace} pair style in tandem with the \texttt{hybrid/overlay} pairstyle to sum the periodic and $\Delta$-MLPs.
All simulations boxes contained 126 waters and were performed at 298 K.
Classical simulations used at 0.5 fs timestep.
The density was obtained from simulations in the isothermal isobaric ensemble (NPT) at a pressure of 1 bar, with a barostat relaxation time of 1 ps.
All density simulations were run for at least 500 ps, with block averaging to obtain the errorbar.
Radial distribution functions and diffusion coefficients were obtained from simulations in the canonical (NVT) ensemble.
For all convergence tests, these were computed at the experimental density, with a box size of 15.577$\,$\AA{} to ensure consistancy.
For the final model production simulations, the RDF and diffusion coefficient were obtained from simulations at the computed density from the NPT simulations.
In all cases, the CSVR \cite{bussiCanonicalSamplingVelocity2007} thermostat was used, with a temperature of 298 K and a temperature relaxation time of 0.1 ps.

\subsection{Simulation details for path integral molecular dynamics}
Nuclear quantum effects were approximately described via ring polymer molecular dynamics simulations, using the i-Pi code coupled with LAMMPS and \texttt{Symmetrix}. For the density, an initial simulation was performed in the NPT ensemble for 400 ps using 32 beads at 298 K and 0.25 fs timestep. A Langevain thermostat with a $\tau$=100 fs, with the PILE thermostat used for the barostat with $\tau$=10 fs. 9 independent replicas were then sampled from this initial simulation and run for between 300 and 400 ps each. The final density was given by the average and standard error over these independent simulations.
To obtain the RDFs and diffusion coefficient, an initial T-RPMD simulation was performed in the NVT ensemble at the above computed density, with 32 beads and a 0.25 fs timestep for 150 ps. A PILE thermostat with $\lambda$=0.5 and $\tau$=100 fs was used. 9 independent configurations were then sampled from this simulation and run for 100 ps each. The diffusion coefficient was then computed from the average of the individual diffusion coefficients of the centroid of the ring polymers as described in further detail in the next section. RDFs were obtained from the positions trajectory of a bead of the ring polymer.

\subsection{Self-Diffusion Coefficients}
The self-diffusion coefficient was obtained from fitting the slope of the mean squared displacement of the water oxygen atoms vs time from 2 to 20 ps. 
The diffusion coefficient obtained for a 126 water simulation box $D(L)$ was corrected for the finite size effects from using a periodic simulation cell of length $L$ using the Yeh and Hummer correction \cite{yehSystemSizeDependenceDiffusion2004}:
\begin{equation}
    D(L) = D(\infty) - \zeta\frac{k_BT}{6\pi \eta L}
\end{equation}
where $\zeta$ was taken as the experimental sheer viscosity of 0.8925 mPas and the numerical coefficient for a cubic simulation cell $\zeta$ as 2.837297.
Table \ref{tab:classical_md} summarises the classical and quantum diffusion coefficients with and without finite size correction:

\begin{table}[h]
\caption{\textbf{Comparison of diffusion coefficient to experiment.} Summary of density and diffusion coefficient  from classical and PIMD simulations at 298 K compared to experiment. Diffusion coefficients show both the finite size corrected value ($D(\infty)$) and before finite size correction ($D(L)$). Error bars are shown in parentheses.}
\label{tab:classical_md}
\begin{tabular}{@{}llll@{}}
\toprule
                 & $\rho$ [g/cm$^3$] & $D(\infty)$ [\AA$^2$/ps] & $D(L)$ [\AA$^2$/ps] \\  \midrule
Experiment       & 0.997                  & 0.23       & -- \\
CCSD(T) (quantum) & 0.989 (0.003)          & 0.22 (0.01)     & 0.18 \\
CCSD(T) (classical) & 0.991 (0.0001)       & 0.21 (0.01)      & 0.17 \\ 
\bottomrule
\end{tabular}
\end{table}

\subsection{Density isobar}
Classical simulations spanning 250K - 330K in increments of 10K were performed to resolve the density isobar at 1 bar.
For each temperature, eight independent replicas were simulated, for at least 3 ns per replica.
Error bars were obtained from the standard error of the independent replicas.

\section{Additional results}

\subsection{Classical results}\label{sec:classical}
\begin{figure}[h!]
    \centering
    \includegraphics[width=1.0\linewidth]{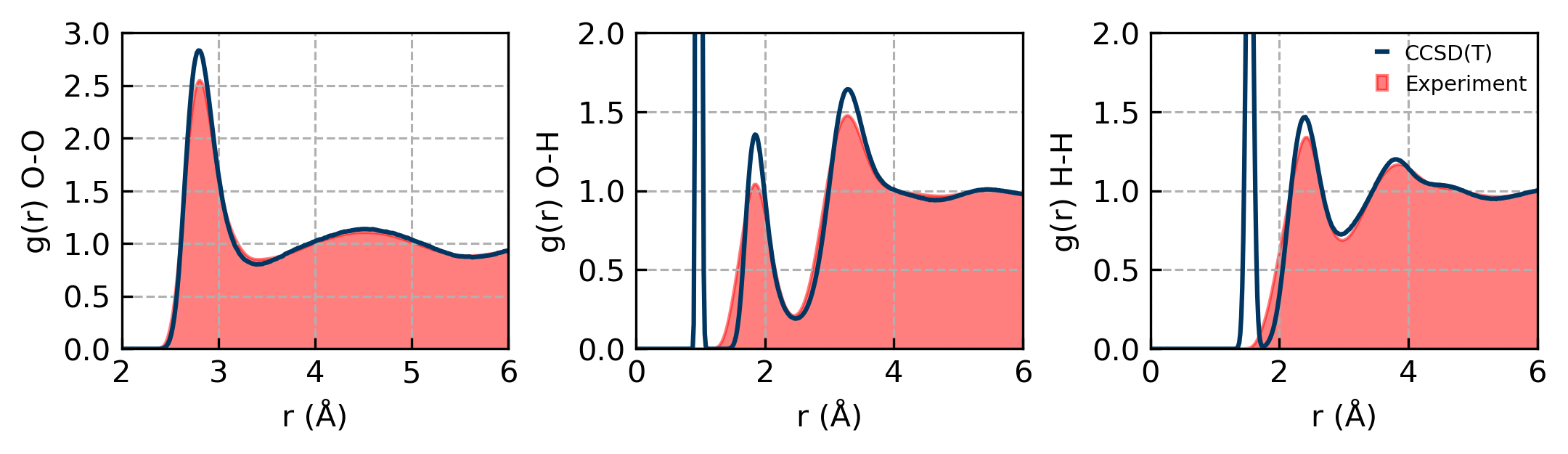}
    \caption{\textbf{Classical CCSD(T) structure of water.} CCSD(T) RDFs for liquid water at 298 K from classical MD simulations compared to experiment}
    \label{fig:rdf_classical}
\end{figure}
Figure \ref{fig:rdf_classical} shows the RDFs obtained from classical MD simulations from the CCSD(T), where the corresponding results including NQEs are given in Figure 3 in the main text.

\subsection{Comparison to literature}\label{sec:comparison-literature}
\begin{figure}
    \centering
    \includegraphics[width=0.5\linewidth]{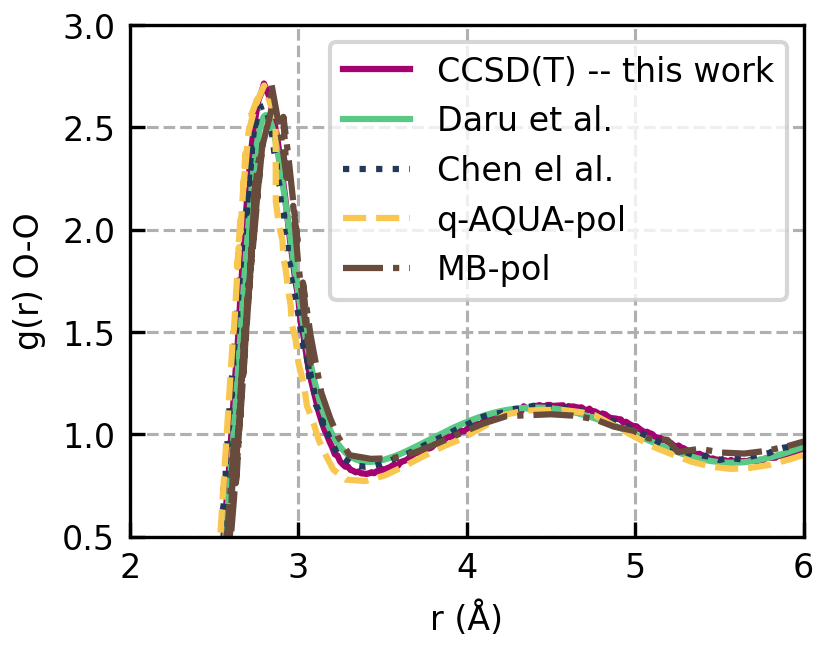}
    \caption{\textbf{Comparison to previous CCSD(T) models.} Comparison of our CCSD(T) O-O RDF from PIMD simulations at 298 K with literature CCSD(T) models. Daru \etal{} comes from PIMD simulations at 298 K from Reference \cite{daruCoupledClusterMolecular2022}. Chen \etal{} comes from PIMD simulations at 300 K from \cite{chenDataEfficientMachineLearning2023}. q-AQUA-pol comes from PIMD simulations at 298 K from Reference \cite{quInterfacingQAQUAPolarizable2023}. MB-pol comes from classical simulations at 298 K from Reference \cite{gartnerAnomaliesLocalStructure2022}. }
    \label{fig:literature-comparison}
\end{figure}

Figure \ref{fig:literature-comparison} compares the O-O RDF predicted by our final $\Delta$-learned CCSD(T) model to previous CCSD(T) models.
We compare to both previous MLP models from Daru \etal{} \cite{daruCoupledClusterMolecular2022} and Chen \etal{} \cite{chenDataEfficientMachineLearning2023}, as well as models based on the many-body expansion, MB-pol \cite{palosCurrentStatusMBpol2024} and q-AQUA-pol \cite{quInterfacingQAQUAPolarizable2023}.
Overall, there is good agreement among all of the models, with some slight differences in the height of the first peak.

\section{Reaching better data efficiency}\label{sec:data_efficiency}

As mentioned in the Discussion of the main text, the total cost to train the final dataset using conservative ORCA settings dataset with the jul-cc-pVQZ basis set is about $1.6\,$million CPUh.
We give the rough breakdown of this total cost for each water cluster radius in Table~\ref{tab:orca_comp_cost}.
The majority of the cost ($1.4\,$million CPUh) comes from the largest cluster radius of $5.5\,$\AA{}.
This consists of 1764 calculations in total, each of which costs roughly 800 CPUh.
As we will show in the next two subsections, both the number of calculations as well as the cost for each cluster can be lowered significantly (by an order of magnitude each), which can substantially lower the cost of this dataset at no loss in accuracy.

\begin{table}[h]
\caption{\label{tab:orca_comp_cost}\textbf{Computational cost of full dataset with TightPNO DLPNO approximation.} A computational cost breakdown for the final dataset when using ``TightPNO'' DLPNO-CCSD(T) in ORCA with the jul-cc-pVQZ basis set. The calculations were performed on a mix of 96-core AMD Genoa nodes and 48-core Intel Cascadelake nodes, with 1,500$\,$GB and 756 $\,$GB or RAM, respectively.}
\begin{tabular}{@{}lrrrrr@{}}
\toprule
Cluster radius (\AA{})                    & 2.5  & 3.5   & 4.5    & 5.5 & Total    \\ \midrule
Total cost (CPUh)                        & 204  & 29910 & 167460 & 1380056 & 1577630 \\
Average number of water molecules & 1.0  & 6.2   & 13.0   & 23.4  &  \\ 
Number of clusters                & 1814 & 1810  & 1810   & 1764 & 7198    \\
Average cost per cluster (CPUh) & 0.1	& 16.5	& 92.5	& 782.3 \\
\bottomrule
\end{tabular}
\end{table}

We have compared our estimates of the costs to that by Daru \etal{}~\cite{daruCoupledClusterMolecular2022}, which has been described~\cite{meszarosShortRangeDMachineLearning2025a} to require ``3,000 DLPNO–CCSD(T) and over 13,000 DLPNO-MP2 for (H2O)\textsubscript{64} clusters''.
It has been described in Ref.~\citenum{daruCoupledClusterMolecular2022} that 109 and 166 days were needed for 10,000 single-point DLPNO-MP2 and DLPNO-CCSD(T) calculations, respectively, using a total of 40 nodes (each with 20 cores).
The resulting cost (in CPUh) for the work of Daru \etal{} is thus 191.5 days on 40 nodes, amounting to ${\sim}3.7\,$ million CPUh.

\subsection{How many 5.5~\AA{} clusters are actually needed?}

\begin{figure}[h]
    \centering
    \includegraphics[width=0.67\linewidth]{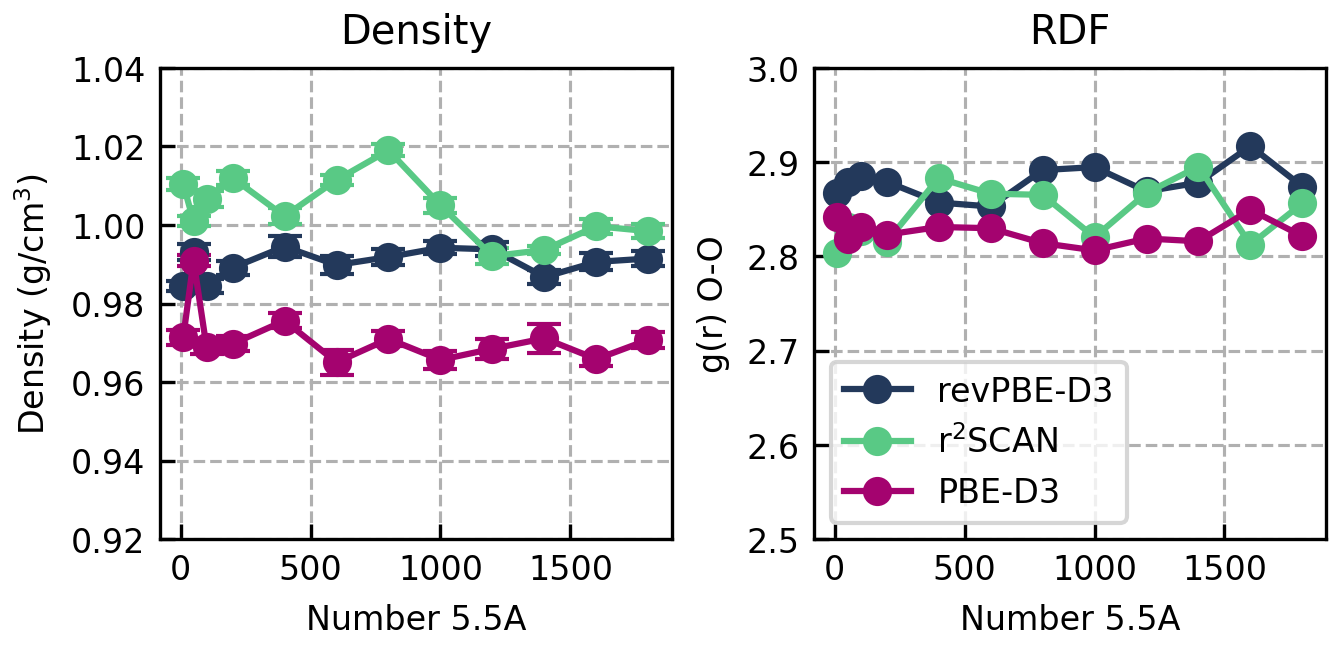}
    \caption{\textbf{Number of 5.5$\,$\AA{} clusters required in dataset.} Comparison of density and RDF for the CCCSD(T) MLP as a function of the number of 5.5$\,$\AA{} clusters added to the total cumulated dataset, with 0 corresponding to the cumulative 4.5$\,$\AA{} dataset. [CCSD(T) level of theory: QZ/TightPNO]}
    \label{fig:cumulative_conv}
\end{figure}

As we will demonstrate in this section, we have generated a relatively conservative dataset for the final CCSD(T) model, and much fewer $5.5\,$\AA{} clusters are actually needed than what we included within our final dataset in Table~\ref{tab:orca_comp_cost}.
In Figure~\ref{fig:cumulative_conv}, we plot the convergence of the final CCSD(T) model for the density and RDF as a function of the number of randomly selected 5.5~\AA{} clusters included within the dataset.
This incorporates all the smaller clusters, so the point at 0 represents the cumulative $4.5\,$\AA{} dataset in Figure~1 of the main text.
We considered three different DFT baselines and it can be seen that beyond $100$ clusters, the density is converged to within $0.01\,$g/cm$^3$ and 0.05 for the PBE-D3 and revPBE-D3 baseline.
This leads to a 5 times overall decrease in computational cost to 275,808$\,$CPUh, as shown in Table~\ref{tab:orca_comp_cost_compact}.
There appears to be more challenges converging the r$^2$SCAN baseline, potentially arising from the lack of long-range dispersion interactions in the functional.

\begin{table}[h]
\caption{\label{tab:orca_comp_cost_compact}\textbf{Computational cost of compact dataset with TightPNO DLPNO approximation.} A computational cost breakdown for the compact dataset when using ``TightPNO'' DLPNO-CCSD(T) in ORCA with the jul-cc-pVQZ basis set. The calculations were performed on a mix of 96-core AMD Genoa nodes and 48-core Intel Cascadelake nodes, with 1,500$\,$GB and 756 $\,$GB or RAM, respectively. The cost for 100 $5.5\,$\AA{} clusters was scaled based on the total cost for the 1764 clusters within the full dataset.}
\begin{tabular}{@{}lrrrrr@{}}
\toprule
Cluster radius (\AA{})                    & 2.5  & 3.5   & 4.5    & 5.5 & Total    \\ \midrule
Total cost (CPUh)                        & 204  & 29910 & 167460 & 78234 & 275808 \\
Number of clusters                & 1814 & 1810  & 1810   & 100 & 5534    \\
\bottomrule
\end{tabular}
\end{table}

\begin{figure}[h]
    \centering
    \includegraphics[width=0.67\linewidth]{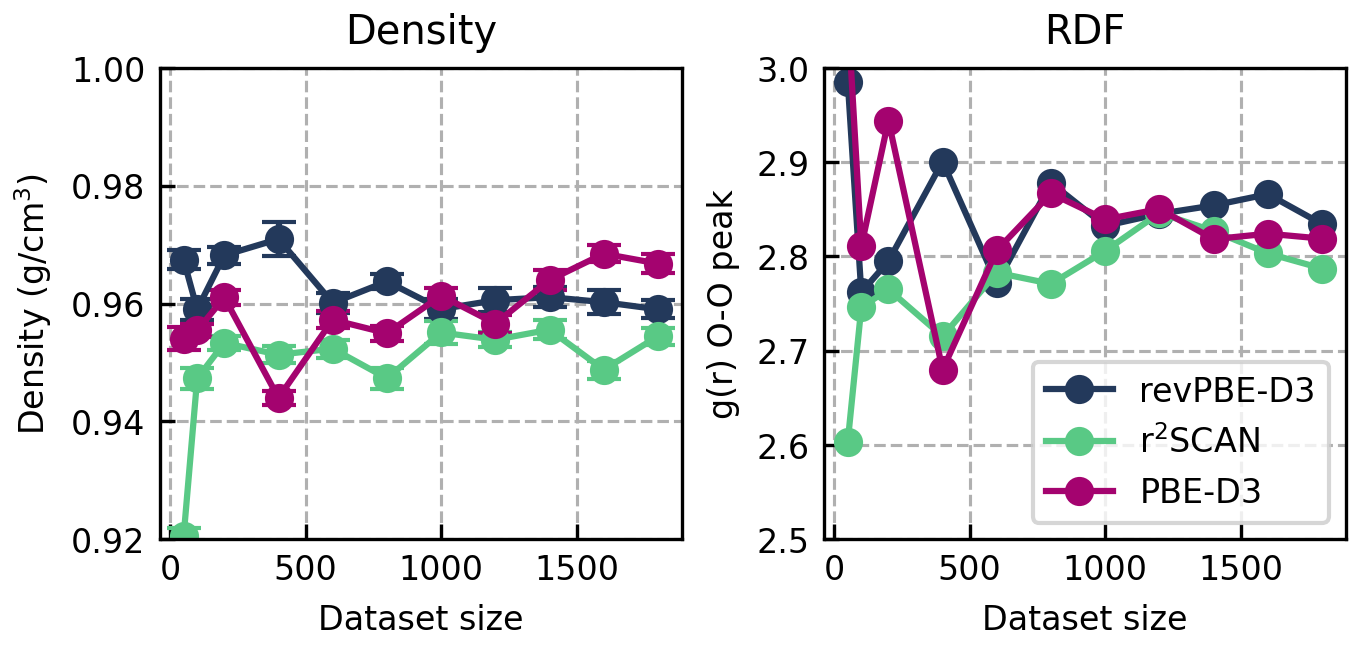}
    \caption{\textbf{Convergence with dataset size of non-cumulative dataset.} Comparison of density, and O-O RDF peak height as a function of total dataset size, where the dataset contains exclusively $5.5\,$\AA{} radius clusters. [CCSD(T) level of theory: QZ/TightPNO]}
    \label{fig:noncumulative_ts_conv_orca}
\end{figure}

It is instructive to highlight that the small number of $5.5\,$\AA{} clusters required to reach convergence arises from our choice of using a cumulative dataset.
For example, in Figure~\ref{fig:noncumulative_ts_conv_orca}, we show the convergence as a function of the number of $5.5\,$\AA{} clusters, where the dataset only include $5.5\,$\AA{} clusters, excluding the smaller clusters from the dataset.
The convergence is less monotonic compared to when the smaller clusters were included in the dataset.
While the density converges to within $0.01\,$g/cm$^3$ within 500 clusters, ${\sim}1500$ clusters were needed to converge both the RDF and diffusion coefficient to $0.01$ and $0.1{\times}10^{-9}$m$^2$/s.

\begin{figure}[h]
    \centering
    \includegraphics[width=0.67\linewidth]{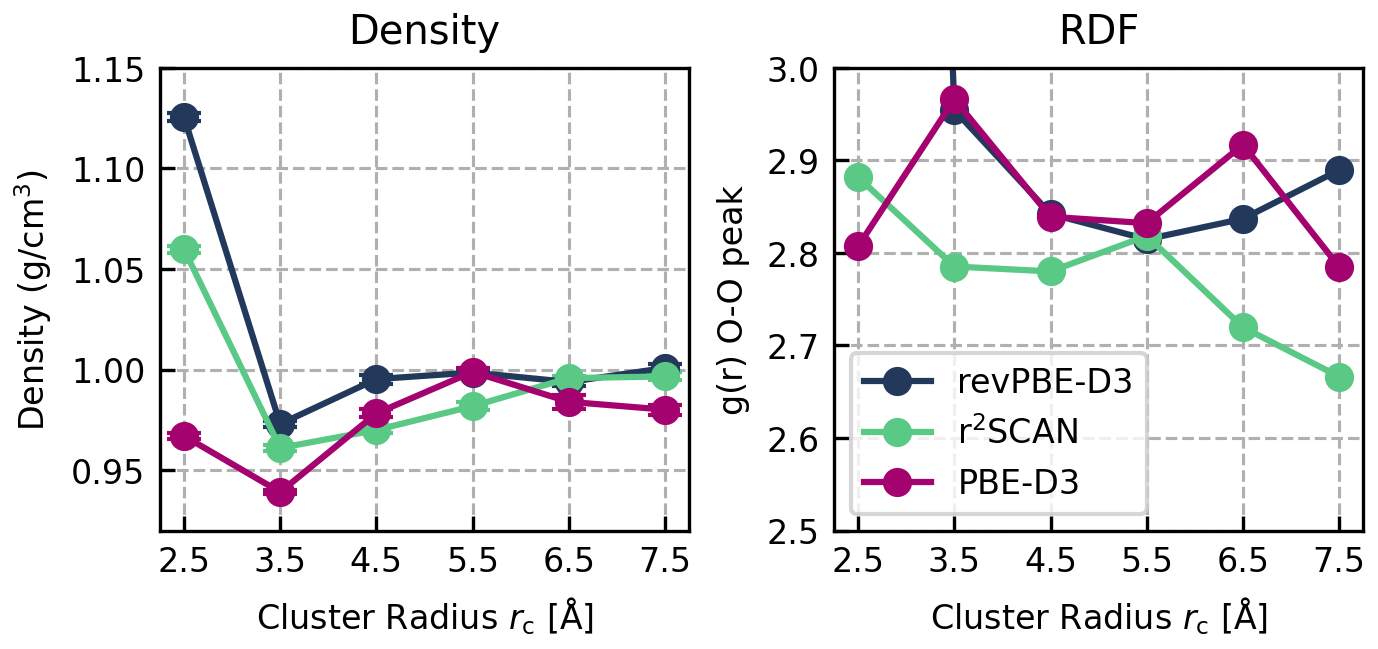}
    \caption{\textbf{Convergence with cluster size of non-cumulative datasets.} Comparison of CCSD(T) density and O-O RDF peak height for models trained on non-cumulative datasets, where a dataset with cluster radius N does not contain clusters of size N-1, N-2 etc. [CCSD(T) level of theory: CBS(DZ/TZ)/NormalLNO]}
    \label{fig:noncumulative_cluster_conv_mrcc}
\end{figure}

Besides a much slower convergence with dataset size, the convergence as a function of cluster size also becomes more challenging using a non-cumulative dataset compared to a cumulative one (see Figure~1 of the main text).
In Figure~\ref{fig:noncumulative_cluster_conv_mrcc}, it can be seen that while the density converges as a function of cluster size, the RDF becomes challenging to converge.
For example, with all three baselines, the RDF peak appears to deivate going towards larger cluster sizes beyond $5.5\,$\AA{}, probably arising because they require more than 1800 clusters to converge with a non-cumulative dataset.

\subsection{Can we make local CCSD(T) cheaper?}

We utilized conservative electronic structure parameters within ORCA as too loose approximations, particularly for the local approximations, can incur noise within the data that can affect the quality of the resulting MLP.
This consists of turning off the RIJCOSX approximation as it was highlighted by Daru \etal{} to bring errors in the forces.
Furthermore, we used the ``TightPNO'' set of thresholds for the DLPNO approximation.
As shown in Table~\ref{tab:orca_comp_cost}, these settings led to an overall cost of $1.6\,$million CPUh.
While this is already less than previous works, it would become more routine if the cost can be decreased further.
In Table~\ref{tab:mrcc_comp_cost}, we show that the total cost for the complete dataset can be lowered by one order of magnitude if we utilize the LNO approximation in MRCC together with less conservative ``normal'' thresholds.

\begin{table}[h]
\caption{\label{tab:mrcc_comp_cost}\textbf{Computational cost of full dataset with ``normal'' LNO approximation.} A computational cost breakdown for the final dataset when using LNO-CCSD(T) in MRCC with the jul-cc-pVQZ basis set and ``normal'' LNO thresholds. The calculations were performed on 48-core Intel Cascadelake nodes with $756\,$GB or RAM.}
\begin{tabular}{@{}lrrrrr@{}}
\toprule
Cluster radius (\AA{})                    & 2.5  & 3.5   & 4.5    & 5.5 & Total    \\ \midrule
Total cost (CPUh)                        & 19 &	1504 &	10318 &	51570 &	63411 \\
Average number of water molecules & 1.0	& 6.2 &	13.0 &	23.6  &  \\ 
Number of clusters                & 1814 &	1814	& 1810 &	1798 &	7236    \\
Average cost per cluster (CPUh) & 0.0 &	0.8	& 5.7 &	28.7 & \\
\bottomrule
\end{tabular}
\end{table}

\subsection{\label{sec:cheapest}What is the most cost efficient dataset?}
\begin{figure}
    \centering
    \includegraphics[width=1.\linewidth]{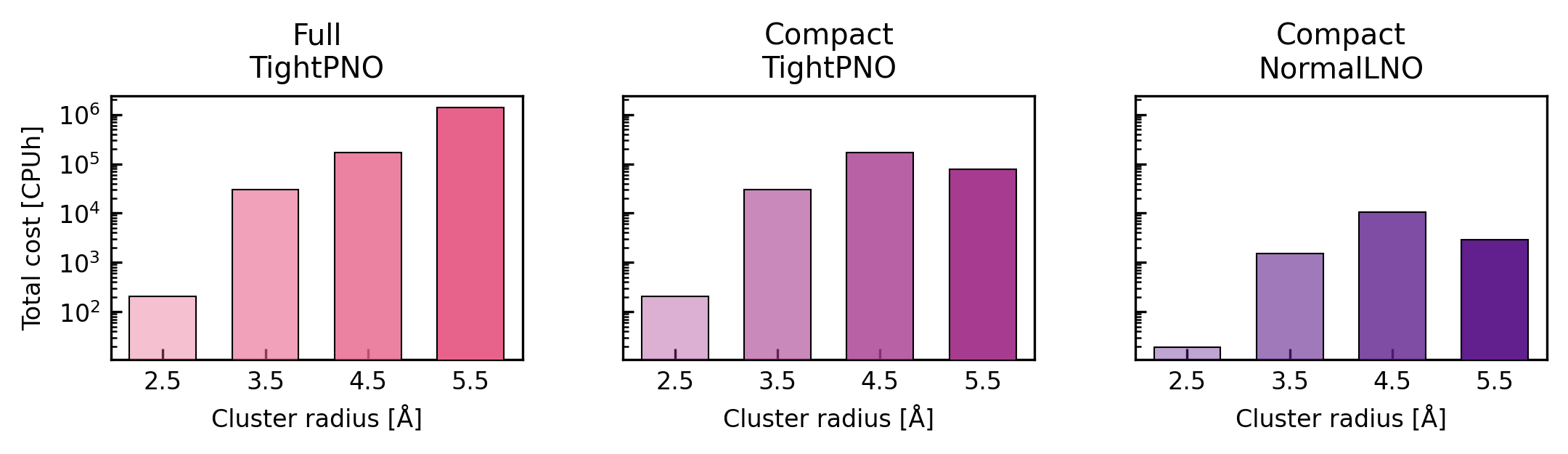}
    \caption{\textbf{Cost breakdown for dataset size and cluster radius.} Breakdown of cost for the total number of each cluster size for the full and compact datasets with both TightPNO and NormalLNO settings.}
    \label{fig:dataset cost}
\end{figure}
\begin{figure}[h]
    \centering
    \includegraphics[width=0.67\linewidth]{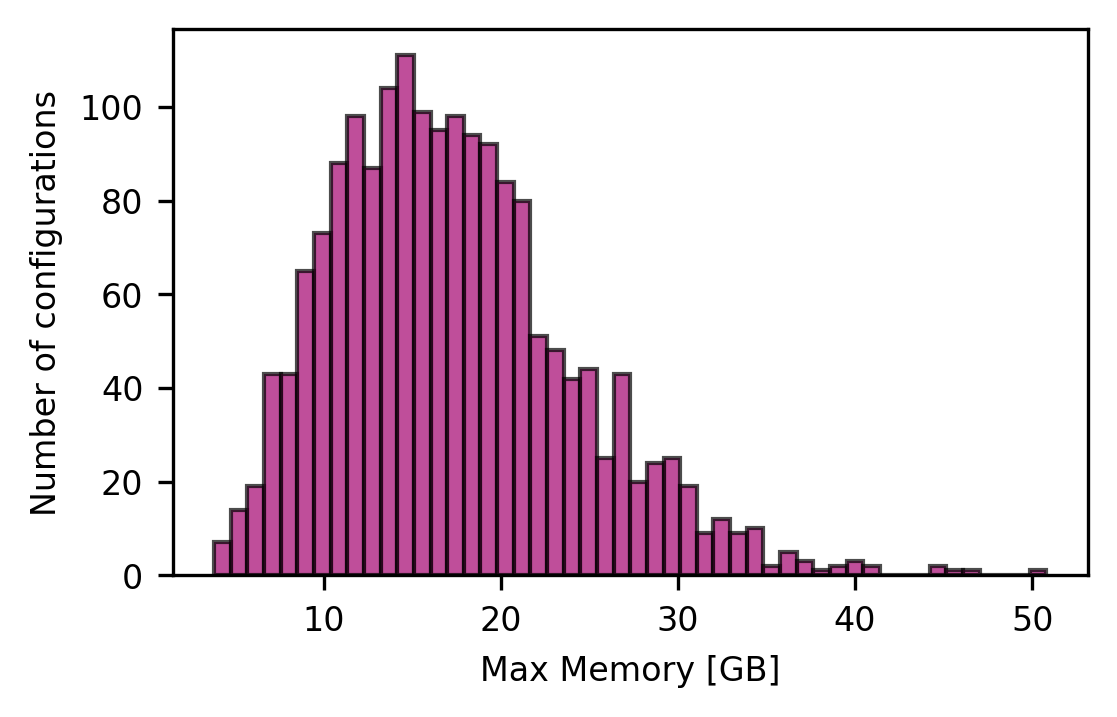}
    \caption{\textbf{Memory usage breakdown.} A histogram of the maximum memory for each $5.5\,$\AA{} cluster calculation using the jul-cc-pVQZ basis set for ``normal'' LNO-CCSD(T).}
    \label{fig:max_mem}
\end{figure}

We consolidate all of the observations made within this section and find that combining the more data-efficient training set together with the less conservative LNO-CCSD(T) settings gives a total dataset cost of $15,000\,$CPUh (see Table~\ref{tab:efficient_comp_cost} and Figure \ref{fig:dataset cost} for a further breakdown).
The resulting cost is more than $100{\times}$ cheaper than the original dataset and electronic structure settings in Table~\ref{tab:orca_comp_cost}.
Furthermore, it is important to also highlight the low computational requirements.
In fact, the efficient dataset can be computed all on a personal (32-core) desktop within the matter of three weeks. 
For example, the average amount of memory -- typically the limiting factor for CCSD(T) calculations -- for the $5.5\,$\AA{} cluster calculations was $17\,$GB, with a maximum of $50\,$GB for the most expensive calculation (see Figure~\ref{fig:max_mem}).
These memory requirements can be easily met with commodity hardware and it is possible to further lower these memory requirements by storing more arrays to the disk using the \texttt{usedisk} parameter within MRCC.

\begin{table}[h]
\caption{\label{tab:efficient_comp_cost}\textbf{Computational cost of compact dataset with ``normal'' LNO approximation.} A computational cost breakdown using the efficient dataset together with LNO-CCSD(T) in MRCC using the jul-cc-pVQZ basis set and ``normal'' LNO thresholds. The calculations were performed on 48-core Intel Cascadelake nodes with $756\,$GB or RAM.}
\begin{tabular}{@{}lrrrrr@{}}
\toprule
Cluster radius (\AA{})                    & 2.5  & 3.5   & 4.5    & 5.5 & Total    \\ \midrule
Total cost (CPUh)                        & 19 &	1504 &	10318 &	2868 & 14710 \\
Number of clusters                & 1814 &	1814	& 1810 &	100 &	5538    \\
\bottomrule
\end{tabular}
\end{table}

